\documentclass{aa}

\usepackage{graphicx,fancyhdr,enumerate}
\usepackage{amsfonts,amsmath,amssymb}
\usepackage{geometry,subfig,xcolor}
\renewcommand\vec[1]{ {\mathbf #1} }% RENEW

\newcommand\cross{\times}

\newcommand\rmd{ {\ \mathrm d} }

\newcommand\deriv[2]{ \frac{\rmd #1}{\rmd #2} }

\def\apjs{\emph{Astrophysical J. Supplement}}

\renewcommand{\vec}[1]{\mathbf{#1}}

\newcommand{\norm}[1]{\left|\left| #1 \right|\right|}

\newcommand{\link}{\mathcal{L}}

\newcommand{\gammat}{\tilde{\gamma}}

\begin{document}

\title{Spatial Scales and Locality of Magnetic Helicity: Part 1} 

\subtitle{} 

\author{C. Prior\inst{1}
  \and G. Hawkes\inst{2}
  \and M.A. Berger\inst{2}}

\offprints{C. Prior, \email{christopher.prior@durham.ac.uk}}

\institute{Department of Mathematical Sciences, Durham University,  Lower Mountjoy, Stockton Rd, Durham DH1 3LE 
  \and Department of Mathematics, University of Exeter, N Park Rd, Exeter EX4 4QF, UK }

\date{Received Day Month 2019 / Accepted Day Month 2019}

\abstract {Magnetic helicity is approximately conserved in resistive MHD models. It quantifies the entanglement of the magnetic field within the plasma. The transport and removal of helicity is crucial in both the dynamo in the solar interior and active region evolution in the solar corona. This transport typically leads to highly inhomogeneous distributions of entanglement.}
{There exists no consistent systematic means of decomposing  helicity  over varying spatial scales and in localised regions. Spectral helicity decompositions can be used in periodic domains and is fruitful for the analysis of homogeneous phenomena. This paper aims to develop methods for analysing the evolution of magnetic field topology in non-homogeneous systems.}{We apply a multiresolution wavelet decomposition to the magnetic field and demonstrate how it can be applied to various quantities associated with magnetic helicity, including the field line helicity. We use a geometrical definition of helicity which allows these quantities to be calculated for fields with arbitrary boundary conditions.}
{It is shown that the multiresolution decomposition of helicity has the crucial property of local additivity and demonstrate a general linear energy-topology conservation law which is a significant generalisation of the two point correlation decomposition used in the analysis of homogeneous turbulence and periodic fields. The  localisation property of the wavelet representation is shown to  characterise inhomogeneous distributions which a Fourier representation cannot.  Using an analytic representation of a resistive braided field relaxation we demonstrate a clear correlation between the variations in energy at various length scales and the variations in helicity at the same spatial scales. Its application to helicity flows in a surface flux transport model show how various contributions to the global helicity input from active region field evolution and polar field development are naturally separated by this representation.}
{The multiresolution wavelet  decomposition can be used to analyse the evolution of helicity in  magnetic fields in a manner which is consistently additive. This method has the advantage over more established spectral methods in that it clearly characterises the inhomogeneous nature of helicity flows where spectral methods cannot. Further its applicability  in aperiodic models significantly increase the range of potential applications. 
%The next logical step is to apply this analysis to actual Resistive MHD simulations which will be the focus of part 2 of this study.
}

\maketitle
\section{Introduction}
The concept of magnetic helicity for a divergence free field $\textbf{B}$  is most commonly introduced as the following scalar integral quantity  
\begin{equation}
\label{eq:1}
H = \int_{\textrm{V}} \textbf{A} \cdot \textbf{B}\  \textrm{dV},
\end{equation}
where $\textbf{A}$ is the vector potential of ${\bf B}$ ($\textbf{B} = \nabla \times \textbf{A}$). This measure was originally introduced by \cite{Woltjer1958}, and given a topological definition by \cite{Moffatt1969} as the linking of magnetic field lines (see also \cite{arnold1998} when field lines do not form closed curves). If we decompose a magnetic field into distinct magnetic regions (by distinct we mean that fieldlines do not cross the boundaries of the regions within the volume $V$) then helicity can be decomposed into the sum of self helicities of each region, and mutual helicities between regions \citep{Berger1999}. For example, if the regions are flux ropes, then the self helicity can be described as the twist and writhe of individual ropes, while the mutual helicity arises from the linking or braiding of distinct ropes. This decomposition has also been applied to the studies of coronal loops: see \cite{Aschwanden2019}, where the authors investigate how stability of coronal loops is associated with the braided linking number. Other shapes are possible: for example an arcade in the solar corona can be sheared (self helicity), but can also envelop a flux rope (mutual helicity).

Magnetic helicity is a well conserved quantity in high resistivity magnetohydrodynamics \citep{Taylor1974,Moffatt2018}. The conservation is maintained in less ideal conditions, albeit to a weaker degree \citep{Berger1984}, making it an ideal approximate invariant for investigation into complex magnetic field systems \citep{ji1995conservation,brandenburg2009critical,contopoulos2009invariant,russell2015evolution,zuccarello2018threshold}).

Magnetic helicity plays an important role in studies of MHD turbulence in general, and dynamo theories of magnetic energy generation in particular (\textit{e.g} \cite{vishniac2001magnetic,Blackman2003,sur2007galactic,brandenburg2017two}). In a two scale kinematic dynamo, the large scale energy can increase exponentially. This poses a problem for magnetic helicity conservation. If the large scale magnetic helicity increases exponentially, then the small scale field must have an equal and opposite helicity which also blows up. Dissipation of the small scale helicity may not be physically feasible. 

A solution to this problem lies in making the dynamo inhomogeneous -- the dynamo operates in one region of space (e.g. the base of the convection zone) and excess magnetic helicity is carried away \citep{brandenburg2009critical,vishniac2013properties}. However, to model this process properly, we need to be able to specify how helicity is spatially distributed. In other words we need to be able to locate where helicity resides more precisely than simply using complete flux ropes.

Another area where helicity localization could be useful is in the study of solar activity. Many studies show how magnetic helicity can flow from the interior into active regions (\textit{e.g.} \cite{berger2000rate,kusano2002measurement,Pevtsov2003,park2008variation,Dalmasse2014,prior2019interpreting}). A knowledge of how this helicity is distributed within the active region may help on the understanding and prediction of flares and coronal mass ejections. Scale dependence of magnetic helicity can also help in understanding the evolution of turbulence in the solar wind (\cite{Brandenburg2011}).

Why is localising helicity so difficult? First, while one can attempt to define helicity density as the quantity
\begin{equation}
H_{den} =  \textbf{A} \cdot \textbf{B},
\end{equation}
this is in no sense gauge invariant as gradient fields can be added to $\mathbf A$ without changing the magnetic field. Second, it does not represent spatially localised information, in a particular gauge such as Coulomb gauge, $\textbf{A}$ is an integral over the magnetic field, and is thus non-local.
This has mathematical grounding -- helicity is associated with the Gauss linking number, for which we must take a double integral across all space. If we only have information about a small patch of field, there is no way of knowing how a field line goes on to twist and writhe around the rest of the field. 

What about integrals of helicity as in equation (\ref{eq:1})?
For the expression (\ref{eq:1}) to be physically meaningful, $V$  must either be unbounded space, or if $V$ is finite with boundary $S$ then $S$ must be a magnetic surface $(\textbf{B} \cdot \hat{\textbf{n}}|_S = 0$). (An alternative is a volume with periodic boundary conditions. However, if there is net flux perpendicular to two periodic directions this can lead to unphysical effects involving magnetic helicity (\cite{Berger1997, Watson2001, Brandenburg2004}).

In section 2 we will review various methods of obtaining localized or semi-localized measures of helicity. First, the relative helicity gives a gauge-invariant measure when the volume is not bounded by a magnetic surface. In general these measures are not additive in the sense that the helicity of all space may not equal the sum of helicities of subvolumes. Next, absolute measures can be found for nested simply connected shells (e.g. concentric spheres). These measures do display additivity. Third,  we discuss Fourier decompositions of helicity, which help to provide information on how helicity behaves on various scales, but without information on locality, this includes a discussion on the transformed two point correlation tensor which decomposes into helicity and energy. Fourth, we discuss field line helicities which measure how one chosen field line interlinks with all other lines. This quantity can be used to accurately quantify reconnection activity in magnetic fields \cite{prior2018quantifying}. The decomposition of helicity into contributions from individual fieldlines is still not localized. Fith and finally, we present helicity densities expressed as two-point correlation functions which are the building blocks of linking and winding. This final measure is gauge invariant, even for fields with non-trivial boundary distributions, as it does not depend on the vector potential for its definition.

Section 3 provides a background to wavelet transforms and multiresolution analysis, and how these can be applied to helicity integrals, and section 4 a formal introduction of the full 3-D wavelet transform and its application to the helicity integral. Section 5 provides examples of how the wavelet multiresolution helicity formulation can be applied in practice. This includes a pair of twisted flux ropes which present a trivial (null) spectral decomposition, the multiresolution helicity decomposition is shown to resolve the spatial separation of the system's entanglement. A second example of a pair of interlinked flux ropes with internal twisting is used to demonstrate how the decomposition can separate out the contributions form large scale linking and smaller scale twisting, as well as correctly asses the localisation of the helicity in this system. In section 6 we consider the application of the multiresolution wavelet decomposition to our geometric two point correlation definition of helicity. This is used to derive linear helicity-energy decompositions for both the helicity and the field line helicity. In section 7 an example of a reconnecting magnetic braid, based on the numerical experiments in \cite{wilmot2009magnetic,wilmot2011heating,russell2015evolution}, is considered. The field line helicity mutiresolution analysis is applied to this system. In particular it is shown that the fields twisted structure and its field line entanglement balance their helicity fluxes at differing spatial scales, and further that the growth then decay in energy of this system at the dominant spatial scales is highly correlated with the field line helicity relaxation at that scale. In section 8 we apply the multiresoltuion decomposition to a flux transport model and finally in section 9 we conclude.	

\section{Helicity measures}

Suppose the volume $V$ is not bounded by a magnetic surface. This introduces a gauge dependence to the helicity integral: given some function $\Phi$ we can let $\textbf{A}_G = \textbf{A} + \nabla\Phi$, which induces a change in helicity corresponding to 
\begin{equation}
H_{G}= H+ \int_{S}\Phi \textbf{B} \cdot d \textbf{S} .
\end{equation}  

\subsection{Relative helicity}
To circumvent this problem, \cite{berger1984topological} introduced a gauge invariant helicity which is referred to as relative helicity. Here, we define the helicity of our magnetic field $\textbf{B}$ within some subvolume of space $V$ relative to a second field ${\bf B}_0$ by taking the difference between the helicity of the fields when we integrate over all space, with the requirement that $\textbf{B}\cdot  \hat{ \textbf{n}} = \textbf{B}_0\cdot \hat{ \textbf{n}}$ on the surface of $V$. This difference is independent of the details of the fields outside $V$. A simple formula for relative helicity is provided by \cite{Finn1985}
\begin{equation}
H_R = \int_{\textrm{V}} (\textbf{A} + \textbf{A}_0)\cdot (\textbf{B} - \textbf{B}_0) \ \textrm{dV}.
\end{equation}
Typically, we take $\textbf{B}_0$ to be a potential field ($\textbf{B}_{0} =  \nabla \psi$), representing the lowest energy field which matches the required distribution upon the surface. 
However, this definition of relative helicity again must be taken over the entire spatial domain: removing the integrand reintroduces a gauge dependence. 
%Additionally, for geometries which are not radially symmetric, we on occasion introduce a current sheet at the boundary of the volume of interest, due to the lack of boundary conditions on the tangential components between $\textbf{B}$ and $\textbf{B}_0$. 
Relative helicity provides a gauge invariant measure of helicity flux \citep{Pariat2005,Dalmasse2014}, which has often been applied to the photospheric field.

Relative helicities can also be defined for other reference fields besides the minimum energy potential field
\citep{park2008variation,valori2012comparing,valori2016magnetic,guo2017magnetic}).
It was shown in \cite{prior2014helicity} that choosing a reference field is essentially equivalent to making a specific choice of gauge. If the reference field is the potential field, and the boundaries of $V$ are planes or spheres then this choice of gauge will be the winding gauge described below. 

\subsection{Absolute helicity} \label{sec:abshel}
An alternate approach which allows for additivity is based on  orthogonal field decompositions, an example in cylindrical geometries is given in \cite{Low2015}. For volumes bounded by arbitrary simply-connected surfaces, an generalised absolute helicity measure is given in \cite{berger2018generalized}. The authors first generalize the toroidal-poloidal decomposition of magnetic fields (e.g. \cite{Moffatt1978}) to geometries without the symmetry of a plane or a sphere. Briefly, $\vec B = {\vec B}_P + {\vec B}_T$ where the poloidal field has no normal current, $\hat n \cdot \vec {J_P} = 0$ and the toroidal field is divergence free and has no normal component, $\hat n \cdot \vec {B_T} = 0$. Similarly, $\vec A = {\vec A}_P + {\vec A}_T$. (Warning: the terminology \emph{toroidal-poloidal} can mean different things in the fusion plasma community (as applied to toruses) compared to the astrophysical plasma community (as applied to spheres)). In asymmetrical volumes, the poloidal field can acquire an extra piece, the \emph{shape field} $\textbf{B}_S$. The authors then define an absolute magnetic helicity 
\begin{equation} \label{eq:abshel}
H_A = \int_{\textrm{V}} \left( 2 \, \vec{A_P} \cdot \vec{B_T} 
 + \vec{A_P} \cdot \vec{B}_S \right) \rm{d}^3   \rm{x}.
\end{equation}
The shape contribution can of course be zero, for instance in the case of a sphere. This helicity can be calculated on successively larger shells, for example concentric spheres; the total helicity within some radius $R$ will be the sum of the helicities of shells from $r=0$ to $r=R$.
On the other hand, if we fill space with an array of cubes, we could calculate $H_A$ for each cube. As the cubes are not nested, additivity will in general not hold.  Of course this approach still has the associated gauge dependence issue.
%In the case of a cube, say, the shape term allows us to properly calculate the helicity captured within any analytical domain: one definition of %spatial helicity could be formed by continuous transformations in scale and shape, although this would not preserve additivity.

\subsection{Fourier spectra}

%In \cite{Brandenburg2005}, the authors define a magnetic helicity density for small scale, turbulent magnetic fields, given by $\textbf{b} = %\textbf{B} - \bar{\textbf{B}}$ where $\bar{ \textbf{B}}$ is a spatial average. 

The splitting of magnetic fields into different scales is core to the study of many magnetohydrodynamical systems: \cite{Verma2004a} provides an in-depth review of turbulent magnetohydrodynamic fields,  which have energy interchanges occurring across a spectrum of spatial scales. 
Following \cite{Blackman2004,Brandenburg2005, Blackman2015}, we can write the magnetic energy spectrum as 
\begin{equation}
E_k = \int |\tilde{ \textbf{B}} |^2 k^2 d\Omega_k, 
\end{equation}
where $k=\norm{\bf k}$, $\Omega_k$ represents a shell in wave space, and the tilde represents the Fourier transform. In Fourier space, we have the direct relation $\tilde{ \textbf{A}}  = -i \textbf{k} \times \tilde{ \textbf{B}}/k^2$. Thus we can write  
\begin{equation}
H_k =  \mathcal{R}e \int   i (\textbf{k} \times \tilde{ \textbf{B}}^{*}(\textbf{k})) \cdot \tilde{ \textbf{B}}(\textbf{k}) d\Omega_k,  
\end{equation}
and as such we have a gauge invariant  measure of magnetic helicity at scale $L = 2\pi /k$ which has the property of additivity, at least is Fourier space (see for example \cite{Moffatt1978,Blackman2003,Demoulin2007,brandenburg2017two}). 

It is important to note that the Fourier decomposition can produce spurious results: if we imagine an infinite system of alternately twisted flux tubes, the Fourier transform of magnetic helicity would be zero at all scales \citep{asgari2009writhe}. To see this, suppose that $B_z$ is constant so only has power at $k=0$, but $B_x$ and $B_y$ vary in $x$ and $y$ to make the oppositely twisted tubes. Then at any $k>0$, both $\vec k$ and $\tilde {\vec B}(\vec k) $ will be in the $x-y$ plane. Thus the triple product must involve $B_z$; but $B_z$ will be zero for $k>0$.

The Fourier spectrum does not give information on locality. The windowed Fourier transform can help. An envelope function with compact support is convolved on top of the infinite sinusoidal functions. Taking the Fourier transform using such a reduced analytic form gives an idea of the variations corresponding to scales at a given locality, but has two downsides (aside from the requirement of periodicity): the transformation does not provide an orthogonal basis, which is required to maintain additivity. Secondly, the window size is fixed, meaning we cannot separate  intense fluctuations which are on smaller scales than the window size from weak contributions on the same scale as the window size.

\subsection{Two point correlation functions}\label{fouriertwopoint}
On a related note the helicity $H_k$ can be related to the magnetic energy $E_k$ via the transform of the two-point correlation tensor $M_{ij}$:
\begin{equation}
M_{ij}({\bf X},{\bf x}) = B_i({\bf X}-{\bf x}) B_j({\bf X}+{\bf x}).
\end{equation}
In a periodic domain one can transform this function over the displacement  ${\bf x}$ to obtain a skew-symmeric tensor function $\tilde{M}_{ij}({\bf X},{\bf k})$ of both postion and scale, and further, for isotropic turbulence, this can be decomposed as
\begin{equation}
\label{helenergyfourier}
\tilde{M}_{ij} = \left[ (\delta_{ij} - k^u_i k^u_j)2 E_k -\mathrm{i}k^u_l\epsilon_{ijl} k H_k \right]/2k^2\Omega_k,
\end{equation}
where $k_i^u$ is the $i^{th}$ component of the unit vector of ${\bf k}$ and $\epsilon_{ijl}$ the alternating tensor
\citep{roberts1975unified,brandenburg2017two}. So the energy is the trace of the tensor $\tilde{M}_{ij}$ and the helicity represented by the off-diagonal components. This is a  potentially powerful relation relating the magnetic helicity and energy on a given Fourier shell at \textbf{each point} of space. If the system is not isotropic (but still periodic) then one can further Fourier transform the tensor over ${\bf X}$ to get a purely spectral tensor $\tilde{M}_{ij}({\bf K},{\bf k})$, then the same relationship (\ref{helenergyfourier}) holds for the Fourier transformed quantities $\tilde{H}$ and $\tilde{E}$ (see \cite{brandenburg2017two} for an application of this decomposition to solar vector magnetogram data). Of course this means we lose the spatial information regarding the energy helicity decomposition which may be crucial for highly inhomogeneous systems.

In this work, we intend to provide a decomposition of magnetic helicity which preserves this additivity and  scale dependence, whilst also providing information about the spatial locality of terms contributing to the power at each scale. Key to our study is the lack of any assumptions about the boundary conditions of the magnetic fields. One result is a variant of (\ref{helenergyfourier}) which can retain information on the spatial distribution of the energy/helicity decomposition even in highly inhomogeneous systems.

\subsection{Geometric definition of helicity: the net winding}\label{sec:winding}
\begin{figure}
\centering
\subfloat[]{\includegraphics[width=3cm]{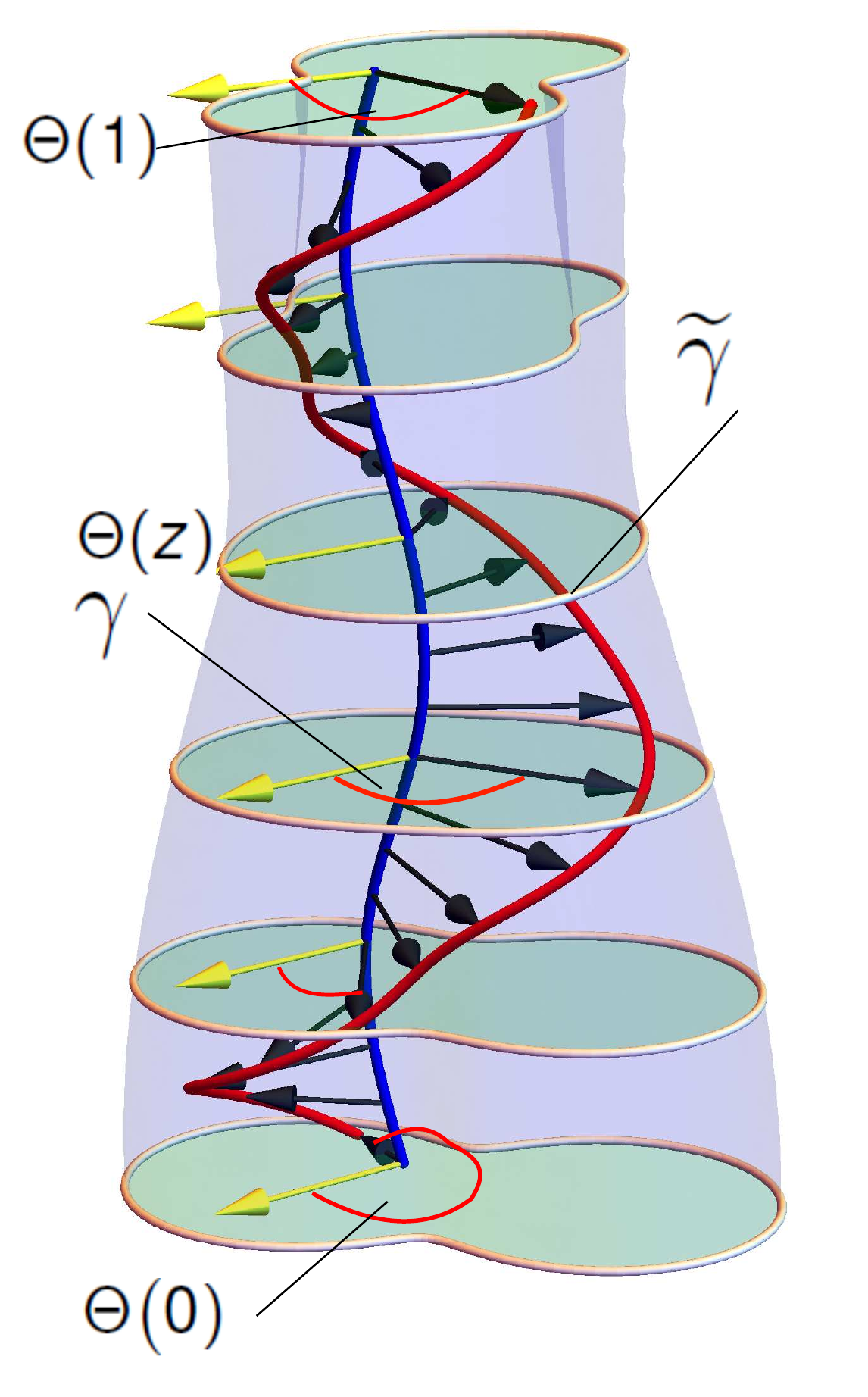}}\quad \subfloat[]{\includegraphics[width=6cm]{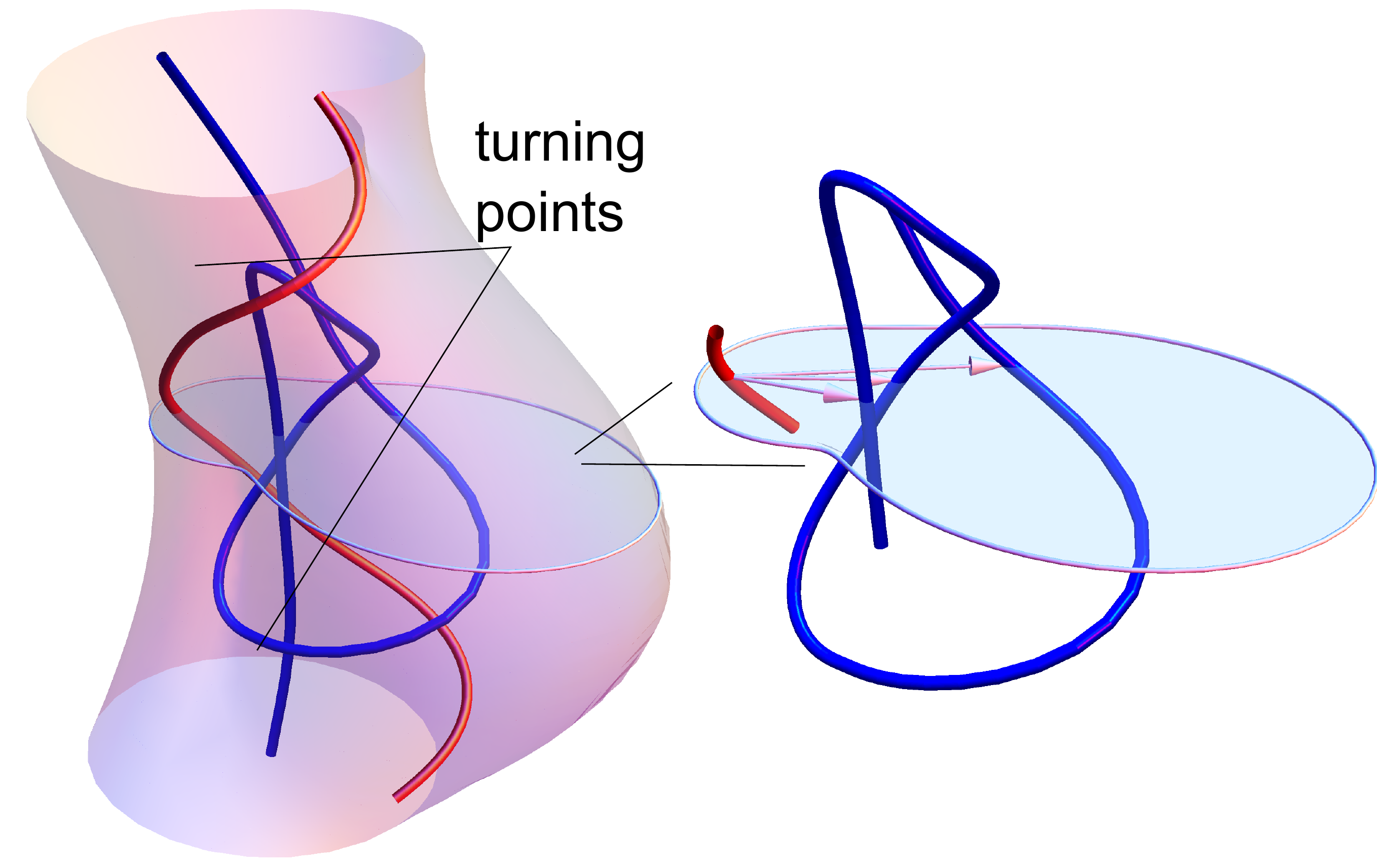}}
\caption{\label{windingfigs} Illustrations of the geometrical interpretation helicity through the winding number. (a) the winding is defined by the mutual angle $\Theta$ between two curves $\gamma$ and $\gammat$. The yellow arrows depict a fixed direction and the black arrows the joining vector of the two curves in a given plane $V_z$ used to define $\Theta$. (b) and example of a pair of curves of which one is non monotonic in $z$. It has two \emph{turning points} at which $\rmd \gamma/\rmd{z}=0$, as shown in figure (b).  This leads to a set of terms which sum to the net winding angle  \ref{eqn:lgg}. } 
\end{figure}

Suppose we consider a volume $\mathcal{V}$ as in figure 1 which can be sliced into parallel planes $z = $ constant. The volume could be infinite in $x$ and $y$, or bounded by a magnetic surface apart from the lower and upper boundaries in $z$. Consider  two field lines $\gamma$, $\gammat$ of ${\bf B}$ that rise monotonically in $z$ (Figure \ref{windingfigs}(a)), so that we can parametrize them as $\gamma=(\gamma_1(z), \gamma_2(z))$, $\gammat=(\gammat_1(z), \gammat_2(z))$. In any planar slice $S_z$, we can use the field line coordinates to define the ``angle'' between the two curves,
\begin{equation}
\Theta(\gamma,\gammat,z) = \arctan\left(\frac{\gamma_2(z) - \gammat_2(z)}{\gamma_1(z) - \gammat_1(z)}\right).
\label{eqn:Theta}
\end{equation}
The net change in this angle as we follow the curves from $z=z_0$ to $z=z_1$ is the pairwise winding number
\begin{align}
\link(\gamma,\gammat)&= \frac{1}{2\pi}\int_{z_0}^{z_1}\frac{\mathrm{d}}{\mathrm{d}z}\Theta(\gamma,\gammat, z)\,\mathrm{d}z \label{eqn:lorig}\\
&= \frac{1}{2\pi}\int_{z_0}^{z_1}\frac{(\gamma_2' - \gammat_2')(\gamma_1 - \gammat_1) - (\gamma_1' - \gammat_1')(\gamma_2 - \gammat_2)}{(\gamma_1-\gammat_1)^2 + (\gamma_2-\gammat_2)^2} \,\mathrm{d}z.
\label{eqn:lgg}
\end{align}
Alternatively, we may write
\begin{equation}
\link(\gamma,\gammat) = \frac{1}{2\pi}\Big(\Theta(\gamma,\gammat, z_1) - \Theta(\gamma,\gammat, z_0)\Big) + N,
\label{eqn:lN}
\end{equation}
where $N$ counts the (signed) number of branch cut crossings  as we follow the curves in $z$. This makes clear that $\link(\gamma,\gammat)$ is invariant under any continuous deformation of the curves that does not move their end-points. In other words, it is a topological invariant.

In \cite{berger2006writhe} it was demonstrated that the definition of $\link(\gamma,\gammat)$ may be generalised to allow for non-monotonic curves (\textit{e.g.} Figure \ref{windingfigs}(b)). Such a curve $\gamma$ is split into $n+1$ sections $\gamma^{(0)}, \ldots, \gamma^{(n)}$ using the $n$ turning points where $\mathrm{d}\gamma_z/\mathrm{d}z=0$. For each section, we define the indicator function
\begin{equation}
\sigma^{(i)} =
\left\{ \begin{array}{cc}
    1 & \mbox{ if } \rmd\gamma^{(i)}_z/\rmd{z}>0,\\
     -1 &\mbox{ if}  \rmd\gamma^{(i)}_z/\rmd{z}<0,\\
      0 &\mbox{ otherwise}.
\end{array}\right.
\label{eqn:sigma}
\end{equation}
We split $\gammat$ and define $\widetilde{\sigma}^{(j)}$ in a similar way. Then the pairwise winding number is the sum
\begin{equation}
\link(\gamma,\gammat) = \sum_{i=0}^n\sum_{j=0}^{\widetilde{n}} \frac{\sigma^{(i)}\widetilde{\sigma}^{(j)}}{2\pi}\int_{z_{ij}^{\rm min}}^{z_{ij}^{\rm max}}\frac{\mathrm{d}}{\mathrm{d}z}\Theta(\gamma^{(i)}, \gammat^{(j)},z)\,\mathrm{d}z,
\label{eqn:l}
\end{equation}
where $[z_{ij}^{\rm min}, z_{ij}^{\rm max}]$ is the mutual range of $z$ values (if any) shared by the curve sections $\gamma^{(i)}$ and $\widetilde\gamma^{(j)}$. Once again, $\link(\gamma,\gammat)$ is invariant to any continuous deformation of the curves that fixes their endpoints. It reduces to (\ref{eqn:lorig}) if both curves have only a single section stretching from $z=z_0$ to $z=z_1$.  Finally in \cite{berger2006writhe} it was also shown that when the curves are closed, $\link(\gamma,\gammat)$ is equal to their Gauss linking integral, hence the notation ``$\link$''.

\begin{figure}
\centering
\includegraphics[width=0.5\textwidth]{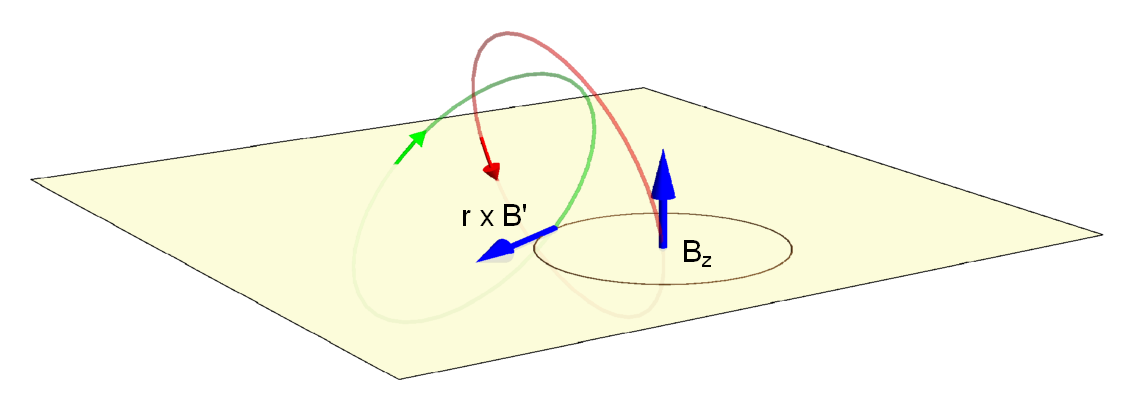}
\caption{\label{fig:twoptcorr} Given two field lines travelling upwards, the product of $B_z$ for the first field line and $\vec r \times \vec B'$ for the second line gives a measure of their mutual winding.}
\end{figure}

For two field lines winding about one another, a particular two point correlation function measures the net winding (see Figure \ref{fig:twoptcorr}). 

\subsection{Fieldline helicity}

Field line helicity is another tool that has become more popular in recent years. For a given field line $\gamma$ we have \citep{Berger1988,yeates2013unique,prior2014helicity,yeates2018relative,moratis2019relative}
\begin{equation}
\mathcal{A}(\gamma) = \int_{{\gamma}} \textbf{A} \cdot \textbf{T} \rmd s,
\end{equation}
where $\textbf{T} = \textbf{B} \||\textbf{B}||$ is the unit tangent vector along the fieldline, and $s$ is the arclength parameter of its curve. The fieldline helicity measures the average winding of all field lines around the field line under analysis, in a similar fashion to the winding gauge developed in \cite{prior2014helicity}. Field line helicity can be seen as the limit of the methodology of \cite{Pevtsov2003}, where each magnetic surface encloses exactly one field line. If we imagine tracing the field lines between two planes, the field line helicity associated with a field line starting at each point $(x_l,y_l)$ on some initial plane (typically taken as the lower boundary of a system) gives a two-dimensional density.

The quantity is not gauge invariant. There is a relative field line helicity version of this quantity, whose definition comes attached with various technical complexities \citep{yeates2018relative,moratis2019relative}, but is an invariant for each individual field line. Further, there is some remaining gauge dependence. 

\subsection{Helicity density as a two point correlation function}

Given any integral representation for $\vec A$, the helicity integral becomes a double integral involving $\vec B$ evaluated at two different points. For example, in Coulomb gauge with $\vec A$ expressed via the Biot-Savart law,
\begin{equation}
H = -\frac{1}{4\pi}\int_{V_z}\int_{V'_z}\frac{{\bf B} \cdot {\bf r}\cross\vec B'}{r^3}\rmd^3 x \rmd^3 x' .
\end{equation}
The integrand can be regarded as a two point correlation function for the magnetic field \citep{Brandenburg2005}.

We will find it useful to employ the winding gauge to give the two point correlation a simple geometrical interpretation.
\cite{prior2014helicity} showed that if one chooses the \emph{winding gauge}  ${\bf A}^{w}$,
\begin{align}
{\bf A}^{w}(x,y,z)=  \frac{1}{2\pi}\int_{V_z}\frac{{\bf B}(x',y',z)\times{\bf r}}{r^2}\, \rmd x'\rmd y'\\
\nonumber {\bf r} = (x-x',y-y',0).
\end{align}
then the helicity is just the flux-weighted average winding of all pairs of field line of ${\bf B}$ with each other, \textit{i.e.}
\begin{align}
\nonumber H({\bf B})= \frac{1}{2\pi}\int_{z_0}^{z_1}\int_{V_z}&\int_{V_z '}B_z({\bf x})B_z({\bf x'})\\\
& \deriv{}{z}\Theta({\bf x},{\bf x'})\rmd^2 {\bf x} \rmd^2 {\bf x'} \rmd{z}
\label{eqn:windingintegral} 
\end{align}
We remark that this requires, as in section \ref{sec:winding}, that the field can be composed of a set of planar surfaces $\mathcal{V} =\left\{V_z\vert z \in[z_0,z_1]\right\}$ and that if the volume is finite in $x$ or $y$ then the field ${\bf B}$ is tangent on the side surfaces.  \cite{berger2018generalized} showed that this relation can be obtained from a  poloidal-toroidal decomposition and extended it to more general domains which can be constructed from sets of simply connected  surfaces.

Further it was shown in \cite{prior2014helicity} that any other choice of gauge, and hence reference field, gives a helicity measure which is equivalent to choosing to measure the angle $\Theta$ with respect to a varying direction, whose rotation is non physical  in that it does not related to the entanglement of the field itself. This is one crucial reason we choose to fix the gauge choice as the winding gauge, \emph{i.e.} because it provides a reliable and meaningful interpretation of the quantity. 

We now represent the helicity as the two point correlation function
\begin{align}
\nonumber &\deriv{H({\bf B})}{z}\\
&= \frac{1}{2\pi}\int_{V_z}\int_{V'_z}\frac{{\bf B}(x,y,z) \cdot \vec B({x',y',z})\cross{\bf r}}{r^2}\rmd x \rmd y \rmd x' \rmd  y'.
\label{eqn:hdens} 
\end{align}
The integrand consists of a  correlation function for the field in the plane $z=z_c$
\begin{equation}
\label{corrden}
\frac{{\bf B}(x,y,z_c)\cdot \vec B({x',y',z_c})\cross\vec r}{r^2}
\end{equation}
which can be regarded as the helicity density of the winding gauge. One can also show the field-line helcity can be written in terms of this correlation function as:
\begin{align}
\label{flhcorr}
\mathcal{A}(\gamma) &=  \frac{1}{2\pi}\int_{\gamma}\int_{V'_z}\frac{{\bf T} \cdot {\bf B}({x',y',z(s)})\cross {\bf r_{\gamma}} }{r_{\gamma}^2}\rmd{s}, \\
\nonumber{\bf r}_{\gamma} &= (x'-\gamma_x,y'-\gamma_y,0),
\end{align}
\cite{prior2014helicity}. This represents the entanglement of the field line $\gamma$ with the rest of the field. Later we will show that, using a wavelet decomposition of ${\bf B}$, it can be represented as a spatial sum of skew symmetric tensors whose trace give the magnetic energy and off-diagonal elements give the helicity, similar to the two point correlation Fourier transform relationship (\ref{helenergyfourier}).

\subsection{A gauge independent definition of the helicity.}
The crucial point about (\ref{eqn:hdens}) is that it gives a definition of a quantity which is gauge independent as it depends only on the field ${\bf B}$ (as does (\ref{flhcorr}) for the fiedline helicity). To be sure we related it to the helicity via a vector potential in (\ref{eqn:windingintegral}) but the following properties can be ascribed to the quantity purely on the basis of its topological definition in terms of winding rate $\rmd \Theta/\rmd z$:
\begin{enumerate}
\item{It is invariant under ideal evolutions which vanish on the domain boundaries \cite{prior2014helicity}.}
\item{It is approximately conserved for low plasma $\beta$ relaxations \cite{russell2015evolution}.}
\item{The field line helicity density can be used to directly quantify magnetic reconnection \cite{prior2018quantifying}, even for fields with normal boundary components.}
\end{enumerate}
So if we take as our definition of the helicity as integrals of the two point correlation function (\ref{eqn:hdens}), then we have absolved ourselves of the gauge problem. What remains is the localisation problem; the  two point correlation function (\ref{corrden}) implies a correlation between all points of space  (in general). 

When applying our spatial decompositions to these gauge free quantities it will be useful to define the following function 
\begin{equation}
\label{corrfunction}
{\bf C}(x,y,z)= \int_{V_z'} \frac{{\bf B}(x',y',z)\times{\bf r}}{r^2}\rmd x'\rmd y'.
\end{equation}
so that the product ${\bf B}\cdot{\bf C}$ represents the total winding weighted correlation of the field at a point $(x,y)$ in the plane $V_z$ with the whole field in that plane. If the field is tangent on the boundaries of the plane $V_z$ then ${\bf  C}={\bf A}^w$, but as we have dicussed above ${\bf C}$ is a meaningful topological quantity even if this is not the case. Then we have the following gauge free definitions of the helcity and field line helcity 
\begin{align}
\label{helnogauge} H({\bf B}) &= \int_{0}^{h}\int_{V_z}{\bf B}\cdot{\bf C}\rmd x\rmd y \rmd z,\\
\label{flhelnogauge}{\cal A}({\bf B})&= \int_{\gamma}{\bf T}\cdot{\bf C}(\gamma)\rmd{s}.
\end{align}

%%%% This is already covered in section 2 %%%%

\section{Spatial localisation of the helicity.}
\begin{figure*}
\centering
\subfloat[]{\includegraphics[width=0.32\textwidth]{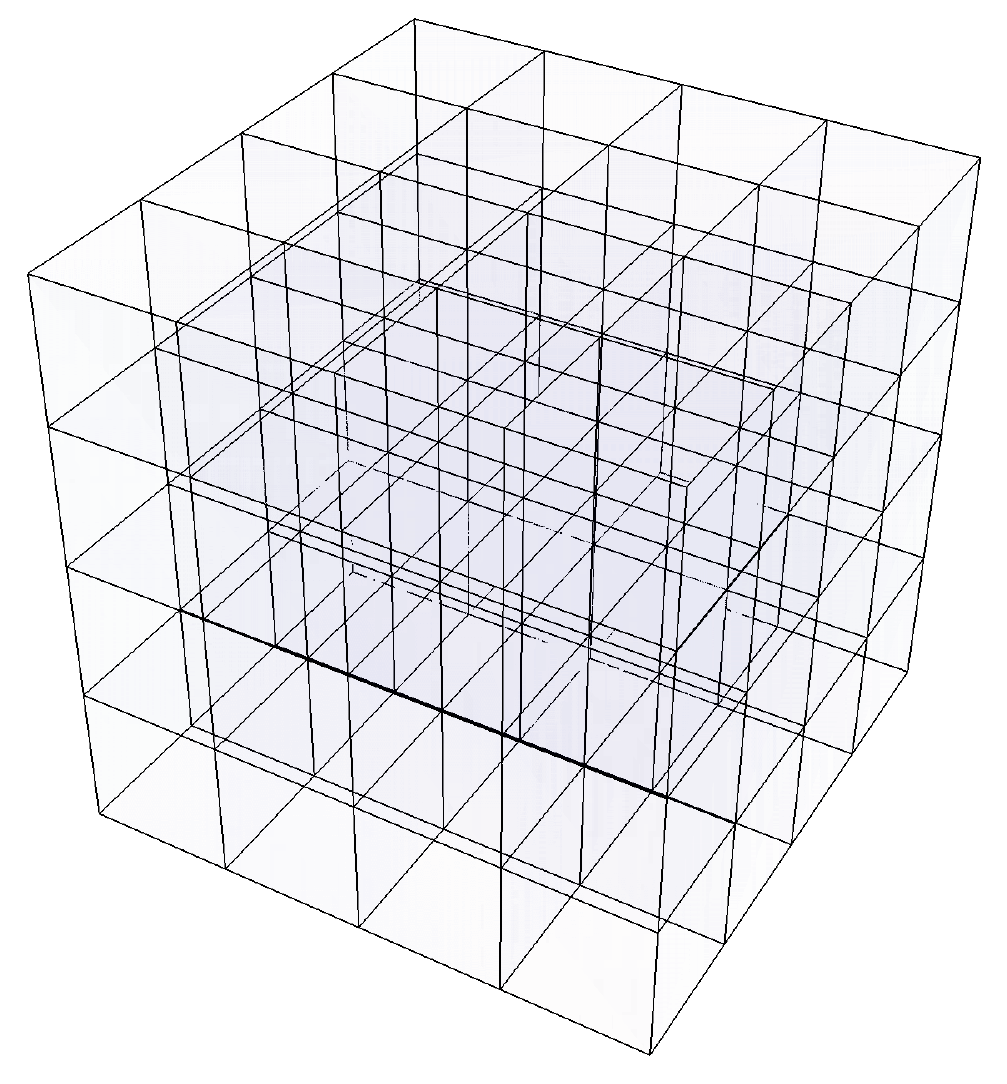}}\quad \subfloat[]{\includegraphics[width=0.32\textwidth]{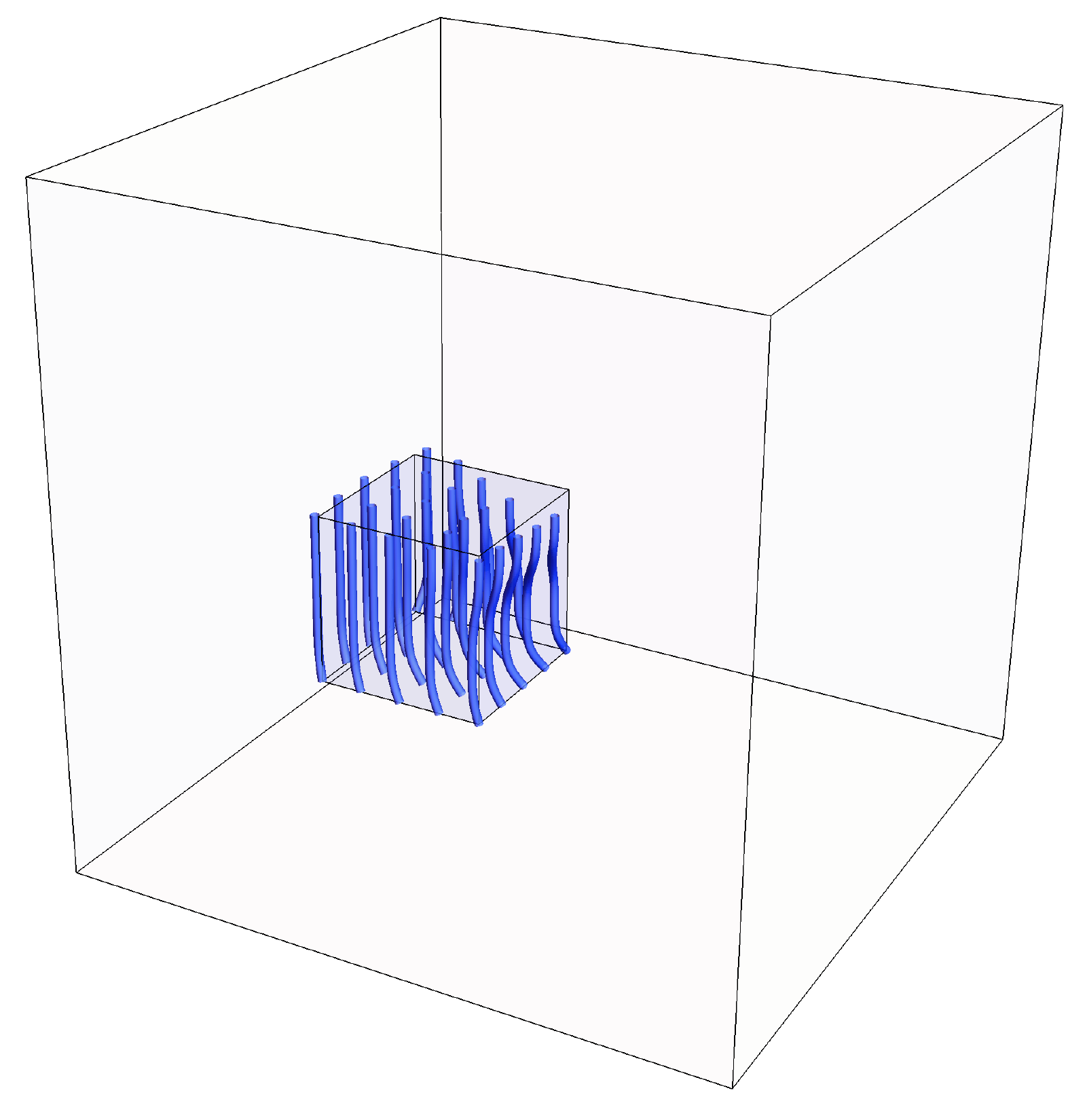}}\quad \subfloat[]{\includegraphics[width=0.32\textwidth]{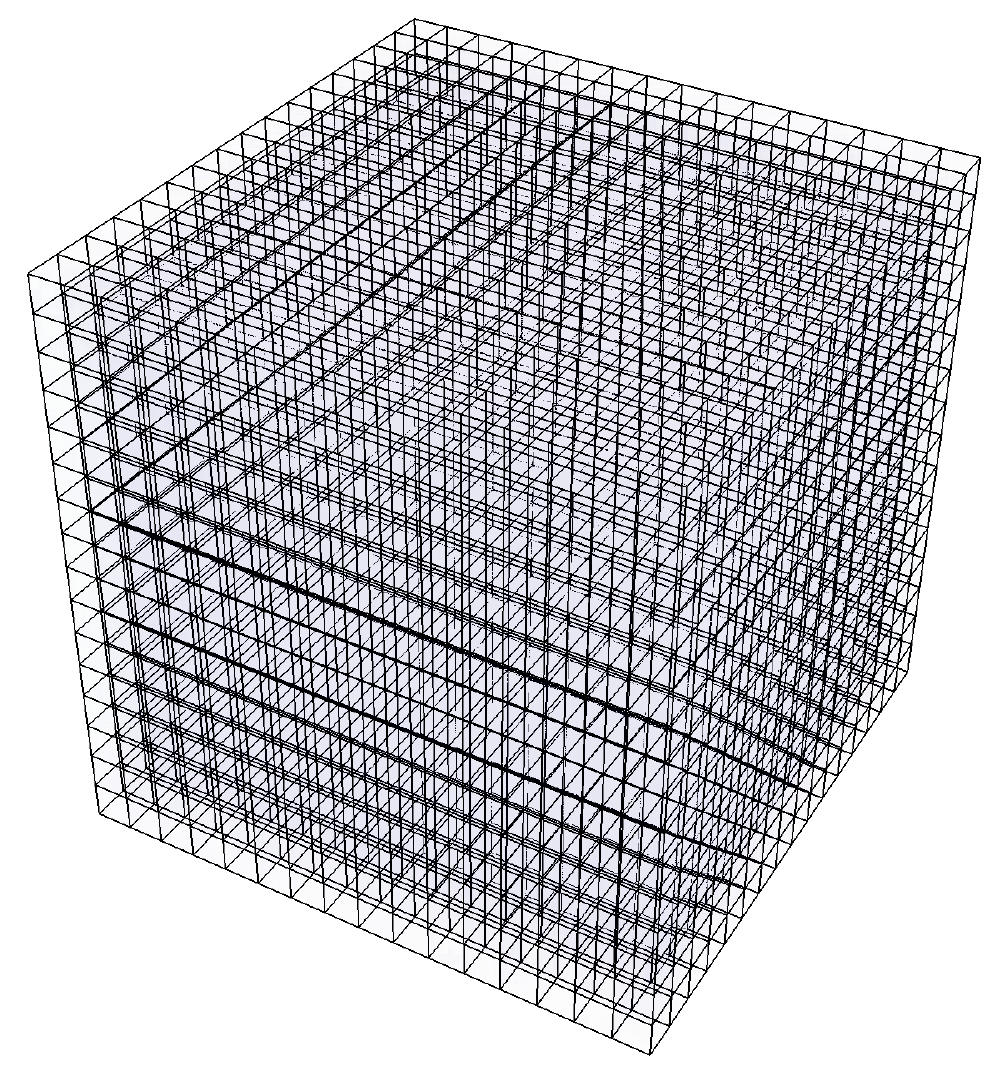}}
\caption{\label{boxfigures} (a) A pictorial representation of splitting a domain into smaller sub-boxes, which each contain a contribution to the total field, as shown in (b). By making the discretisation even more dense, as shown in (c), we make our approximation more accurate.}
\end{figure*}

Here we very briefly introduce the ideas underpinning our spatial localisation technique to give some geometrical intuition as to its interpretation.
In order to spatially decompose the helicity we need a representation of the field ${\bf B}$ in terms of functions  with compact support, for example the box function
\begin{equation}
\label{harr}
\phi(x) = \left\{ 
\begin{array}{cc}
1 & \mbox{ if }  0\leq x\leq 1,\\
0 & \mbox{ if } x>1 \mbox{ or } x<0,
\end{array}
\right.
\end{equation}
whose 3-D composite
\begin{equation}
\Phi_{x_{0}y_{0}z_0}(x_1,x_2,z) =  \phi(x-x_{0})\phi(y-y_{0})\phi(z-z_0),
\end{equation}
gives a 3-D box of compact support which is translated in space. By discretizing the domain \textit{i.e.} $x_{0}= i\Delta x,\,i \in 1,\dots N$, a set of non-overlapping box functions $\Phi_{lmn}$ could be created to cover the domain in a non overlapping fashion (\textit{e.g.} Figure \ref{boxfigures}(a)). One could then approximate the mean field ${\bf B}$ as
\begin{align}
\label{bapprox}
&{\bf B}(x_1,x_2,z) \approx  \sum_{l=1}^{N}\sum_{m=1}^{N}\sum_{n=1}^{N}{\bf B}_{lmn}\Phi_{lmn}(x_1,x_2,z), \\
&\quad {\bf B}_{lmn}= \int_{V}{\bf B}\,\Phi_{lmn}\rmd{V}.
\end{align}
Each coefficient $B_{lmn}$ would be representative of the average behaviour of the field in the box $(l,m,n)$ ( Figure \ref{boxfigures}(b)).

One could do something similar for the vector potential (if we use the winding gauge we obtain the two point correlation definition of the helicity). Using the fact the function $\Phi_{lmn}$ has compact support an approximation to magnetic helicity would then be given by the sum
\begin{equation}
\label{hdecomp}
H({\bf B}) \approx   \sum_{i=1}^{N}\sum_{j=1}^{N}\sum_{k=1}^{N}{\bf B}_{lmn}\cdot {\bf A}_{lmn},
\end{equation}
(one could think of this as a spatial decomposition of the constant part of the Fourier series), each triplet $(lmn)$ would give the average of the density ${\bf A}^w\cdot{\bf B}$ in a particular cube of the domain.
To capture the local variation, we could use a function such as
\begin{equation}
\label{harrscale}
\phi(x) = \left\{ 
\begin{array}{cc}
1 & \mbox{ if }  1/2\leq x\leq 1/2,\\
-1 & \mbox{ if } 0<x\leq 1,\\
0 & \mbox{ if }\mbox{ if } x>1 \mbox{ or } x<0,
\end{array}
\right.
\end{equation}
the coefficients of which could then be added to (\ref{bapprox}) to give a more accurate approximation of the field (this is a little like breaking the sinusoid of the Fourier transform into sub components). The smaller the discretization size (the spatial scale) $N$ the more accurate the approximation (the discretization in Figure \ref{boxfigures}(c) would be more accurate than that in (a)).

Of course there are multiple issues with such a decomposition. For example how do we choose the scale of decomposition? In fact (with regards to the varying component) we might want to choose multiple scales for fields which have both large and small scale variation. How might we then add up these scales whilst avoiding redundancy? A branch of wavelet analysis called Multiresolution analysis tells us exactly how to perform such a decomposition orthogonally, and combine it across multiple scales. We shall introduce it formally in Section \ref{sec:wavelets}: the localised functions used above are the so-called Harr Wavelet (\ref{harr}) and Harr scaling function (\ref{harrscale})) and the sum (\ref{bapprox}) forms one part of the decomposition; the method for combining the varying field components is a little more complex. We will see in Section \ref{sec:3dexamples} that this spatial scale decomposition (multiresolution analysis) of the helicity for the two example fields discussed above show a non trivial (absolute) variation across spatial scales which naturally identifies to ``size" and position of the helicity producing components of the field.

\begin{figure*}
\centering
\subfloat[]{\includegraphics[width=0.32\textwidth]{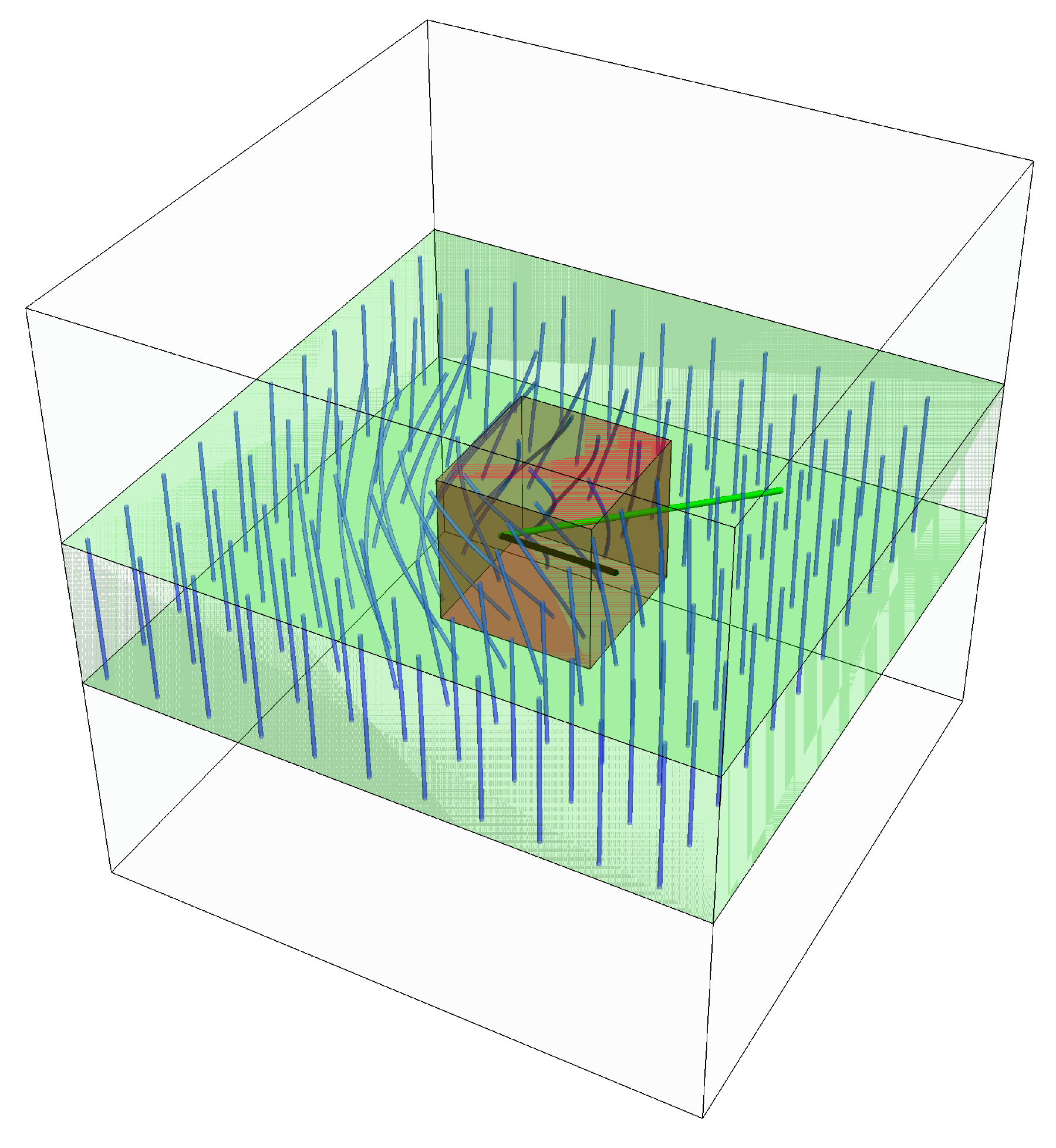}}\quad \subfloat[]{\includegraphics[width=0.32\textwidth]{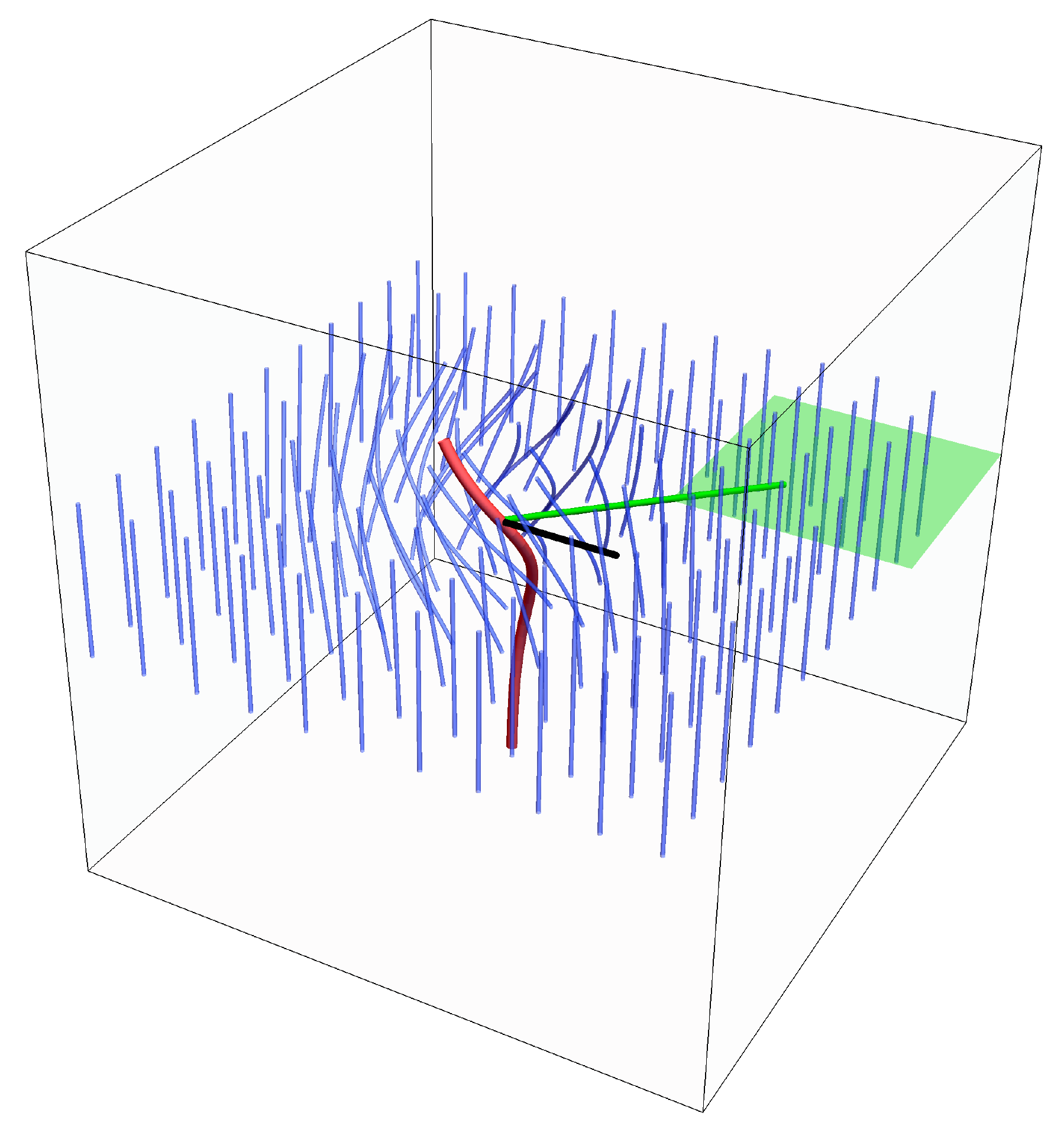}}
\caption{\label{fieldlinehel}Illustrations of the various geometric interpretations of helicity calculations involving the winding gauge ${\bf A}^w$/two point correlation function (\ref{corrden}). (a) The geometrical interpretation of the spatial contribution ${\bf A}^w_{lmn}\cdot{\bf B}_{lmn}$ of a spatial (wavelet) decomposition of the helicity. The red box represents the spatial sub-domain given by the triplet $lmn$. Each point in this red domain contributes a winding with the rest of the field in the plane in which it is contained. Because ${\bf A}^w_{lmn}\cdot{\bf B}_{lmn}$ is a sum over the whole red domain, the set of planes containing the red domain provide winding contributions to the sum, as indicated in the figure. (b) The winding contribution obtained by spatially decomposing the field ${\bf B}$ \textbf{inside} the two point correlation function. It represents the winding of the red curve $\gamma$ (localised in the plane as represented by the vector ${\bf B}$ at that point) and the field lines in the sub-plane indicated in green.}
\end{figure*}

\subsection{Two point winding correlation localisation}

A decomposition such as (\ref{hdecomp}) is still not fully localised, since the integral ${\bf A}^w$ integrates across planes of constant $z$ of the domain. The quantity ${\bf A}^w_{lmn}\cdot{\bf B}_{lmn}$ actually represents the winding correlation of the field in the box of compact support with the rest of the field in the planes containing that box, as indicated in Figure \ref{fieldlinehel}(a). A finer localisation can found be inserting the decomposition of the field ${\bf B}$ into the two point correlation function which will lead to terms in the form
\begin{equation}
{\bf B}({\bf x})\cdot\frac{{\bf B}_{lm}\Phi_{lm}\times{\bf r}}{r^2}\rmd^2 y
\end{equation}
 where the field decomposition is for a specific $z$ value hence we would use a two dimensional decomposition,\ textit{i.e.}
\begin{align}
\label{bapprox2d}
&{\bf B}(x,y,z) \approx  \sum_{i=1}^{N}\sum_{j=1}^{N}{\bf B}_{lm}(z)\Phi_{lm}(x,y) \\
&\quad {\bf B}_{lm}= \int_{V_z}{\bf B}(x,y,z)\,\Phi_{lm}\rmd x\,\rmd y.
\end{align}
This would yield a local density representing the winding of the curve at the point $(x,y)$ with a localised sub domain of curves, as indicated Figure \ref{fieldlinehel}(b)). We use this decomposition to develop a spatial  helicity energy decomposition for both the helicity and field  line helicity in section (\ref{sec:helicityenergy}).

\section{Helicity,Wavelets and Multiresolution Analysis} \label{sec:wavelets}
We first consider dimensional scalar functions $f(x)$, whose 3-D wavelet representations are composed of combinations of one dimensional wavelets. We will focus on the set of wavelets known as discrete wavelets which form the discrete wavelet transform, in particular the multiresolution representation of this transform (for details on the continuous wavelet transform see \textit{e.g.}  \cite{grossmann1990reading}). What follows is far from a comprehensive overview of the mathematics the multiresolution analysis which is beyond the scope of this study, for a more detailed introduction see \textit{e.g.} \cite{farge1992wavelet} for a practical introduction in a fluid dynamic context and  \cite{jawerth1994overview} for more details  on the underlying mathematics.
\subsection{The basic idea for Harr wavelets} 
\begin{table*}[!h]
\begin{center}\
 \begin{tabular}{||c c||}
 \hline
 Scale & Intervals (Locality) \\ [0.5ex] 
 \hline
 0 &  $[0,1]$\\ 
\\
 1 & $[0, \frac{1}{2}]$, $[\frac{1}{2}, 1]$\\
\\
 2 & $[0, \frac{1}{4}]$, $[\frac{1}{4}, \frac{1}{2}]$, $[\frac{1}{2}, \frac{3}{4}]$, $[\frac{3}{4}, 1]$\\
\\
 3 & $[0, \frac{1}{8}]$, $[\frac{1}{8}, \frac{1}{4}]$, $[\frac{1}{4}, \frac{3}{8}]$, $[\frac{3}{8}, \frac{1}{2}]$, $[\frac{1}{2}, \frac{5}{8}]$, $[\frac{5}{8}, \frac{3}{4}]$, $[\frac{3}{4}, \frac{7}{8}]$, $[\frac{7}{8}, 1]$\\
 \\
 
 \hline
 \end{tabular}
 \caption{Illustrative Examples of Scales and Locality}
 \label{tab:scales}
 \end{center}
\end{table*}
In the previous section we introduced the the varying Harr Wavelet (\ref{harr}) and mean (scaling) function (\ref{harrscale}) used to characterise the varying and mean behaviour of a function $f(x)$ respectively, over a given subset of the domain. The basic idea of a multiresolution analysis for some discrete signal on a domain $[0,1]$ (one can always scale this to any finite domain) is that we can choose the domain spatial scales as factors of two \textit{i.e.} $2^s,\dots s\in 1,2,....$ as indicated in Table \ref{tab:scales}. For a given choice of scale $s$ the functions (\ref{harrscale}) and (\ref{harr}) mutually are orthogonal and orthonormal with each other if suitably dilated and translated \textit{i.e.},
\begin{align}
 &\int_{0}^{1}\sqrt{2^{s}}\phi(2^s x-i)\sqrt{2^{s}}\psi(2^s x-j)\rmd{x} =0,\\
\nonumber & \int_{0}^{1}\sqrt{2^{s}}\phi(2^s x-i)\sqrt{2^{s}}\phi(2^s x-j)\rmd{x}  \\
 & =\int_{0}^{1}\sqrt{2^{s}}\psi(2^s x-i)\sqrt{2^{s}}\psi(2^s x-j)\rmd{x} =\delta_{ij}.
\end{align}
A common notation for these dilation/translation combinations is to write
\begin{equation}
\phi_{si}(x) = 2^{s/2}\phi\left(2^s x-i \right) \mbox{ and } \psi_{si}(x) = 2^{s/2}\phi\left(2^s x-i \right)
\end{equation}
so that 
\begin{align} 
&\nonumber  \int_{0}^{1}\phi_{si}(x)\psi_{sj}(x)\rmd{x} = 0, \\
& \int_{0}^{1}\phi_{si}(x)\phi_{sj}(x)\rmd{x} = \int_{0}^{1}\psi_{si}(x)\psi_{sj}\rmd{x} =\delta_{ij}.
\label{orthog2}
\end{align}
One can also see some of these conditions can be extended for comparisons between scales,
\begin{align}
\nonumber\int_{0}^{1} \phi_{si}(x) \psi_{s'j}(x)&=0,\forall s'\geq s \mbox{ and } \\
 & i \in 0,1,\dots 2^{s}-1,j\in 0,1,\dots 2^{s'}-1,
\label{orthog3}
\end{align}
as well as 
\begin{align}
\label{orthog4}&\int_{0}^{1} \psi_{si}(x) \psi_{s'i'}(x)=\delta_{ss'ii'}.
\end{align}
Thus if we pick some base scale $s_b$ the orthogonality conditions (\ref{orthog2}), (\ref{orthog3}) and (\ref{orthog4}) ensure it is possible to write
\begin{align}
\nonumber f(x) &= \sum_{i=0}^{2^{s_b}-1}\left<\phi_{s_bi}\vert f\right>\phi_{s_b i}(x)  \\
& +\sum_{s=s_b}^{\infty}\sum_{i=0}^{2^{s}-1}\left<\psi_{si}\vert f\right>\psi_{si}(x) , 
\label{eq:multires}
\end{align}
where
\begin{equation}
\label{normdef}
\quad \left<g,f\right> = \int_{0}^{1}f g \rmd{x}.
\end{equation}
for square integrable functions on $[0,1]$ \cite{jawerth1994overview}. It is a celebrated result of Ingrid Daubechies \cite{daubechies1993wavelet} to demonstrate that there are various classes of functions $\phi$ and $\psi$ with compact support which satisfy the conditions (\ref{orthog2}, \ref{orthog3}, \ref{orthog4}), such that (\ref{eq:multires}) can be used to represent square integrable functions. The specific choice of $\phi$ and $\psi$ can often be quite important (for discussions on the matter see \textit{e.g.} \cite{farge1996wavelets,zhang2004wavelet}). The example calculations detailed in this study were performed with various wavelet choices, but no significant differences were observed, so these comparative calculations were omitted for brevity. In what follows all example calculations use the Harr  basis.
 
In practice the series (\ref{eq:multires}) will be truncated at some maximum scale $s_m$ and the most common choice is to have $s_b=0$, which prioritises the number of spatial scales used in the expansion, so that we have the following multiresolution approximation:
\begin{equation}
\label{multires3}
f(x) \approx  \left<\phi_{0}\vert f\right>\phi_{0}(x) +\sum_{s=0}^{s_m}\sum_{i=0}^{2^{s}-1}\left<\psi_{si},f\right>\psi_{si}(x).
\end{equation}
In what follows we use the equality sign for series such as (\ref{multires3}) on the assumption it is understood this is actually an approximation.

\subsection{Three dimensional multi-resolution analysis}
In a three-dimensional Cartesian domain $V$, we expand the behaviour along each direction via a one dimensional multi-resolution expansion \textit{i.e.} we assume a 3-D function $H({\bf x})$ can be written as $H_x(x)H_y(y)H_z(z)$ (\cite{jawerth1994overview}). By writing each function as a multi-resolution expansion we will encounter $8$ types of combinations ($4$ in 2-D) for each scale $s$:
\begin{equation}
\label{eq:dirs}
  \psi^{\mu} _{slmn} (\textbf{x})=\begin{cases}
    \phi_{sl}(x) \phi_{sm}(y) \phi_{sn}(z) & \text{if $\mu = 1$},\\
    \psi_{sl}(x) \phi_{sm}(y) \phi_{sn}(z) & \text{if $\mu = 2$},\\
    \phi_{sl}(x) \psi_{sm}(y) \phi_{sn}(z) & \text{if $\mu = 3$},\\
    \phi_{sl}(x) \phi_{sm}(y) \psi_{sn}(z) & \text{if $\mu = 4$},\\
    \psi_{sl}(x) \phi_{sm}(y) \psi_{sn}(z) & \text{if $\mu = 5$},\\
    \psi_{sl}(x) \psi_{sm}(y) \phi_{sn}(z) & \text{if $\mu = 6$},\\
    \phi_{sl}(x) \psi_{sm}(y) \psi_{sn}(z) & \text{if $\mu = 7$},\\
    \psi_{sl}(x) \psi_{sm}(y) \psi_{sn}(z) & \text{if $\mu = 8$}.\\
    
  \end{cases}
\end{equation}
Writing the respective coefficients as 
\begin{equation}
    H_{slmn}^{\mu} =\int_\textbf{V} H(\textbf{x})\psi^{\ \mu}_{slmn}(\textbf{x})   \ d\textbf{x},
\label{eq:coeff}
\end{equation}
the ensuing multi-resolution decomposition will be 
\begin{align}
\nonumber  H(\textbf{x}) &= H^{1}_0 \psi_{0}({\bf x}) \\
   &  +\sum_{s=0}^{s_m} \sum_{l=0}^{2^{s}-1}\sum_{m=0}^{2^{s}-1}\sum_{n=0}^{2^{s}-1} \sum_{\mu = 2}^{8} H_{slmn}^{\mu} \psi^{\mu}_{slmn}(\textbf{x}),
   \label{multiresseriesfull}
\end{align}
see \textit{e.g.} \cite{farge1996wavelets}.
\subsubsection{Compacted notation}
Throughout this study we will not be paying particular attention to the contributions of individual $\mu$ terms, thus for each $l,m,n$ we shall assume the $\mu$ summation has been performed. Further, for brevity we define an index $ k$ which, when summed over will be assumed to indicate a sum over $l$, $m$ and $n$ (or just $l$ and $n$ in 2-D). Thus we write (\ref{multiresseriesfull}) as
\begin{equation}
\label{multiresseries}
    H(\textbf{x}) = \sum_{s=0}^{s_m} \sum_{k=0}^{2^{s}-1}H_{sk}\psi_{sk}(\textbf{x}).
\end{equation}

\subsubsection{Relative scale contributions}
For a function $H$ one can define the total (relative) contribution $C_{s}(H)$ to the multiresolution decomposition at a scale $s$ as
%\begin{equation}
%%C_{S}(H) = \frac{\sum_{k=0}^{2^s-1}H_{sk}}{\sum_{s=0}^{s_m}\sum_{k=0}^{2^s-1}H_{sk}}.
%\end{equation}
\begin{equation}
C_{s}(H) = \sum_{k=0}^{2^s-1}H_{sk}.
\end{equation}
Similarly we define the relative power $P_s(H)$ at scale $s$ as:
\begin{equation}
P_{s}(H) = \frac{\sum_{k=0}^{2^s-1}\vert H_{sk} \vert }{\sum_{s=0}^{s_m}\sum_{k=0}^{2^s-1}\vert H_{sk}\vert}.
\end{equation}
For comparison, if the function has the required periodicity we can calculate the power contained at each Fourier scale $k$ through the quantity
\begin{equation}
 H_k= \sum _{|\textbf{k}| =k} H_k(\textbf{k}).
\end{equation}
where the $H_k$ are the coefficients of the appropriate Fourier series of $H$.
\subsection{Helicity formulae} \label{sec:wavelethelformula}
We consider multi-resolution expansions  (\ref{multiresseries}) for ${\bf B}$ (one per component) and  the  multiresolution expansion of ${\bf A}$, \textit{i.e.}
\begin{equation}
{\bf A}=  \sum_{s=0}^{s_m} \sum_{k=0}^{2^{s}-1}{\bf A}_{sk} \psi_{sk}(\textbf{x}),
\end{equation}
and substitute them into the the helicity integral $\int_{V}{\bf A}\cdot{\bf B}\rmd V$. Using the orthogonality relationships (\ref{orthog2}), (\ref{orthog3}) and (\ref{orthog4})  we obtain a summation over the coefficients of the two series
\begin{equation}
\label{helsum}
H= \sum_{s=0}^{s_m}\sum_{k=0}^{2^{s}-1}{\bf A}_{sk}\cdot{\bf B}_{sk}= \sum_{s=0}^{s_m}\sum_{k=0}^{2^{s}-1}H_{sk}.
\end{equation}
So $H_{sk}$ is the  helicity contribution at scale $s$ at position $k=lmn$ (summing over all directions $\mu$). As indicated in Figure \ref{fieldlinehel}(a), if ${\bf A}$ is the winding gauge, then the geometrical interpretation of this coefficient corresponds to the winding of the compact domain of scale $s$, centered at the coordinates indicated by the triplet $k$, with the $z$-slice of the volume containing  this domain. 

\subsection{Absolute Helicity}
Here we perform a multiresolution analysis of absolute helicity (introduced in Section \ref{sec:abshel}), which has a physical interpretation similar to that of the winding gauge. The expression introduced in Section \ref{sec:abshel} (equation \ref{eq:abshel}) was suitable for a series of infinite Cartesian planes (with $\textbf{B}_{S} = 0$), but for finite domains we must take account of boundary terms. This can be done by properly expanding the sum
\begin{align}
\nonumber H_A &=  \int_V (\textbf{A}_{\textbf{P}} + \textbf{A}_{\textbf{T}}) \cdot (\textbf{B}_{\textbf{P}} + \textbf{B}_{\textbf{T}}) \rmd^3 \textbf{x}, \\
& =\int_V \left( \, \vec{A_P} \cdot \vec{B_T} + \vec{A_P} \cdot \vec{B_P} + \vec{A_T} \cdot \vec{B_P}   \right) \rm{d}^3   \textbf{x},
\end{align}
($\textbf{A}_{\textbf{T}}\cdot \textbf{B}_{\textbf{T}} = 0$) and as such
\begin{align}
\nonumber H_A = \sum_{s=0}^{s_m}\sum_{k=0}^{2^{s}-1}&\bigg[ \vec{A}_{\textbf{P},sk} \cdot \vec{B}_{\textbf{T},sk} + \vec{A}_{\textbf{P},sk} \cdot \vec{B}_{\textbf{P},sk} \\
&  + \vec{A}_{\textbf{T},sk} \cdot \vec{B}_{\textbf{P},sk}   \bigg],
\end{align}
following our notation.
\section{Multiresolution Analysis of Magnetic Helicity: Illustrative Examples} \label{sec:3dexamples}
In this section we present two examples which illustrate the benefits of the spatial decomposition offered by a multiresolution analysis of magnetic helicity. Unless otherwise stated, all quantities in this section have arbitrary units, and are calculated within the winding gauge which for both fields is equivalent to the absolute helicity representation
\subsection{Oppositely Twisted Flux Tubes} \label{sec:opptwisttubes}
\begin{figure}[h]
	\begin{center}

      \includegraphics[width=0.5\textwidth]{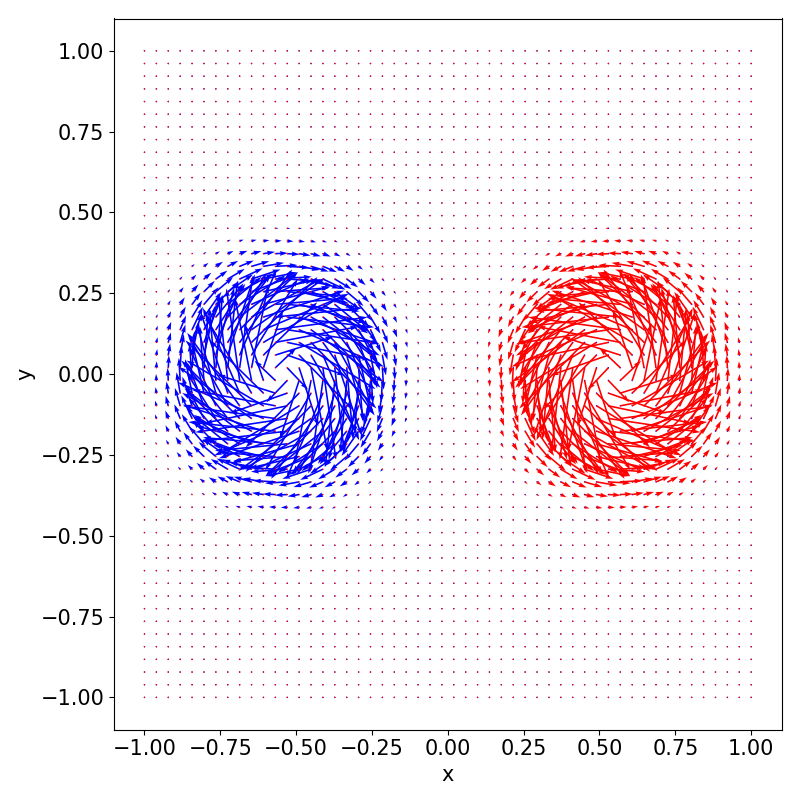}
      \caption{Magnetic field vector plot of equation (\ref{eq:opptwisteq}) at $z = 0$, red indicates positive twist, and blue indicates negative twist.}
      \label{fig:opptwistfluxtubes}
    \end{center}
\end{figure}
\begin{figure}[h]
	\begin{center}
      \includegraphics[width=0.5\textwidth]{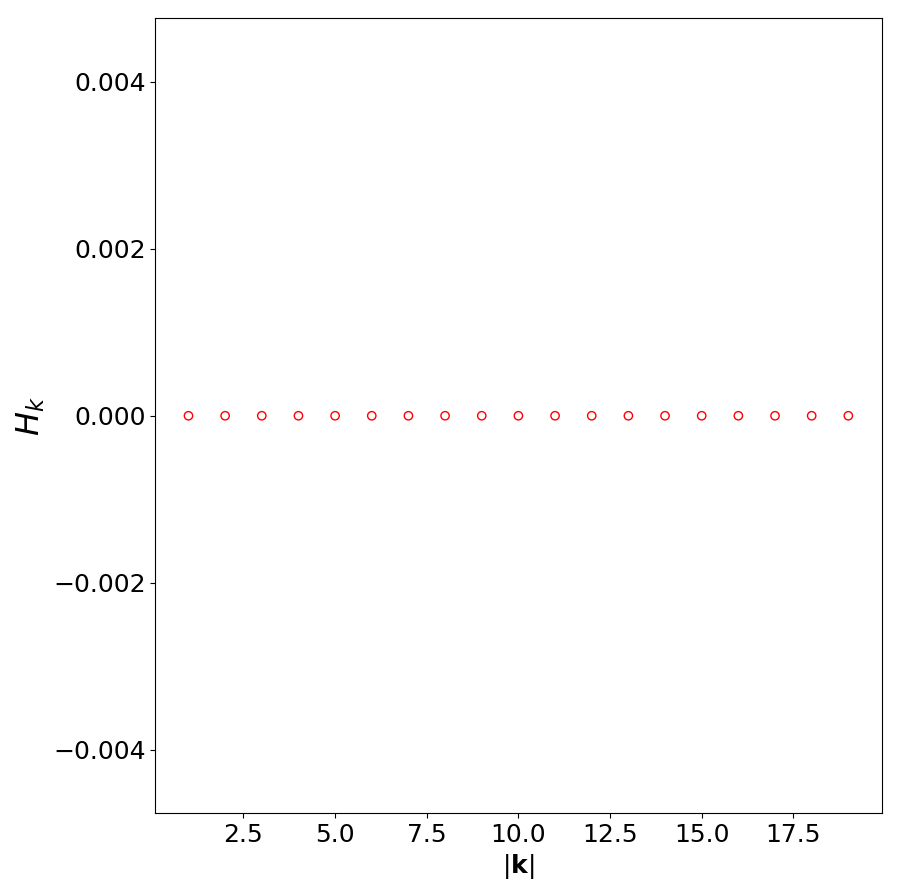}
      \caption{Plot of fourier decomposition $H_k$ of magnetic helicity determined by equation (\ref{eq:opptwisteq}).}
      \label{fig:opptwistfourier}
    \end{center}
\end{figure}
The first of our illustrative examples is that of a pair of oppositely twisted flux tubes whose vector potential takes the following form:
\begin{align}
\textbf{A} &= (-5y, 5x,  50a\exp\bigg[\frac{ -((x+ 0.55)^2 + y^2)}{a^2}\bigg] \nonumber \\
& -  50a\exp\bigg[\frac{ -((x- 0.55)^2 + y^2)}{a^2}\bigg]),
\label{eq:opptwisteq}
\end{align}
where we have taken $a = 0.2$. This field (independent of $z$) is visualised for the domain $[-1,1]\times[-1,1]$, in Figure \ref{fig:opptwistfluxtubes}.
Making an assumption of periodicity (which can be interpreted as an infinite repeating pattern of the form shown here), Fourier analysis indicates that that this magnetic field has zero overall helicity at every scale, even when the sum over ${\bf k}=k$ is taken with absolute values, as shown in Figure \ref{fig:opptwistfourier}.

\begin{figure*}[h]
    \centering
    \includegraphics[width=\textwidth]{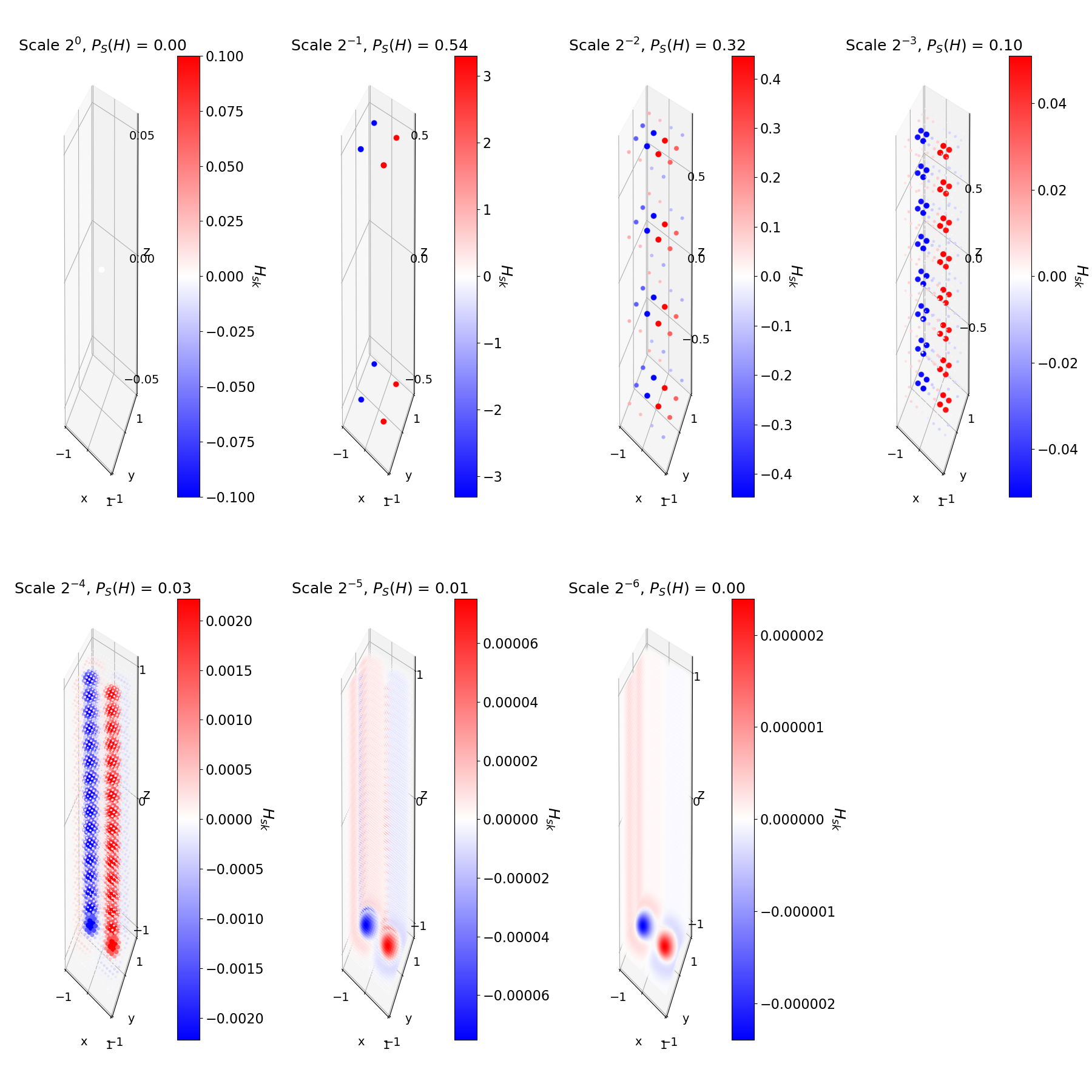}
    \caption{$H_{sk}$ for $s = 0 \rightarrow 6$ associated with the magnetic field distribution in equation (\ref{eq:opptwisteq}). The largest scale is a measure of the numerical round-off. At the two smallest scales $2^{-5, 6}$, the visual appearance of the bubblegram is distorted by the frequency of data points. }
    \label{fig:opptwisthelbubble}
\end{figure*}

By contrast, in Figure \ref{fig:opptwisthelbubble}, we plot the $H_{sk}$  for the wavelet multiresolution analysis of the magnetic helicity at spatial scales $r = 0 \rightarrow 6$ (along with the associated power $P_s(H)$). The plotting style is that of a "bubblegram": each sub-domain of helicity given by the multiresolution analysis is allocated a three-dimensional sphere at its centre. The radius of this sphere is dependent upon the absolute magnitude of the helicity of the sub-domain, and its color, red or blue, indicating a positive/negative sign respectively. 

\begin{figure}[!h]

    \centering
    \includegraphics[width=0.5\textwidth]{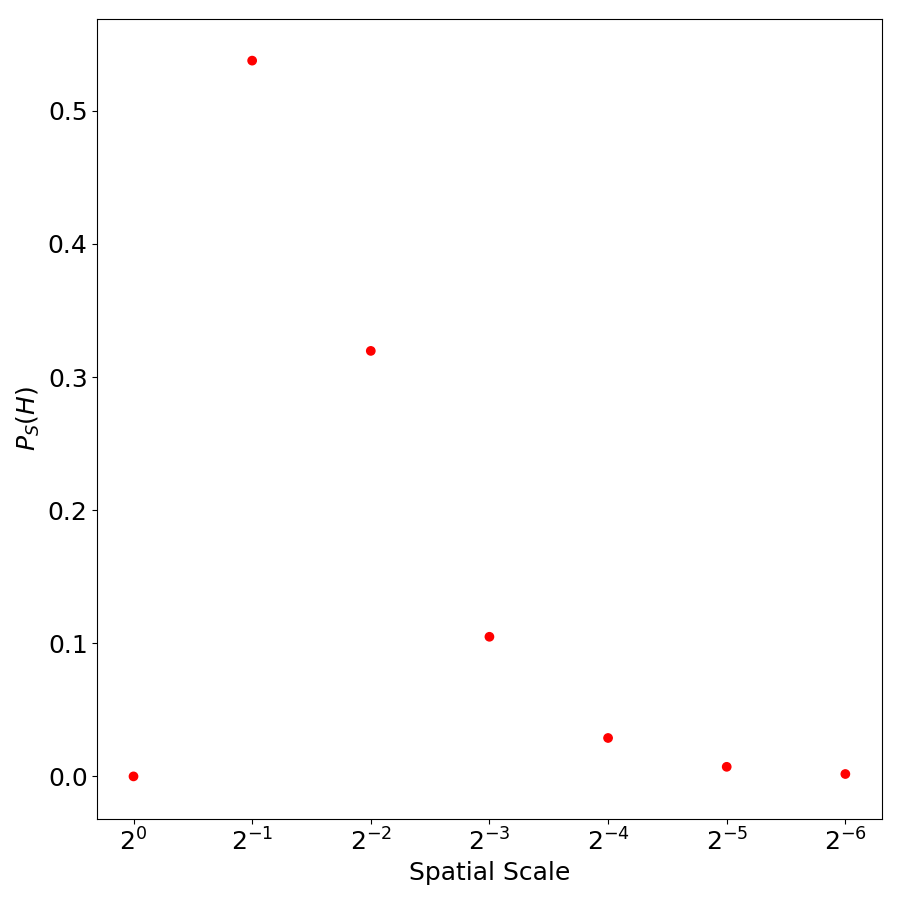}
    \caption{$P_s(H)$ for the multiresolution helicity expansion of the field (\ref{eq:opptwisteq}) at all spatial scales.} 
    \label{fig:opptwistabsscale}
\end{figure}

The bubblegrams indicate that the helicity is well localised in space in accordance with Figure \ref{fig:opptwistfluxtubes}, presenting with the correct sign of twist. It can be seen that the total helicity $C_s(H)$ at each scale is zero. The absolute magnetic helicity power $P_s(H)$ is well localised in scale, as indicated in Figure \ref{fig:opptwistabsscale}. Peak magnetic helicity occurs at half the spatial scale of the domain, which is in good agreement with the distribution of the twist in the magnetic field itself. 

\subsection{Linked Rings} \label{sec:rings}
\begin{figure}[h]
	\begin{center}
 \includegraphics[width=0.5\textwidth]{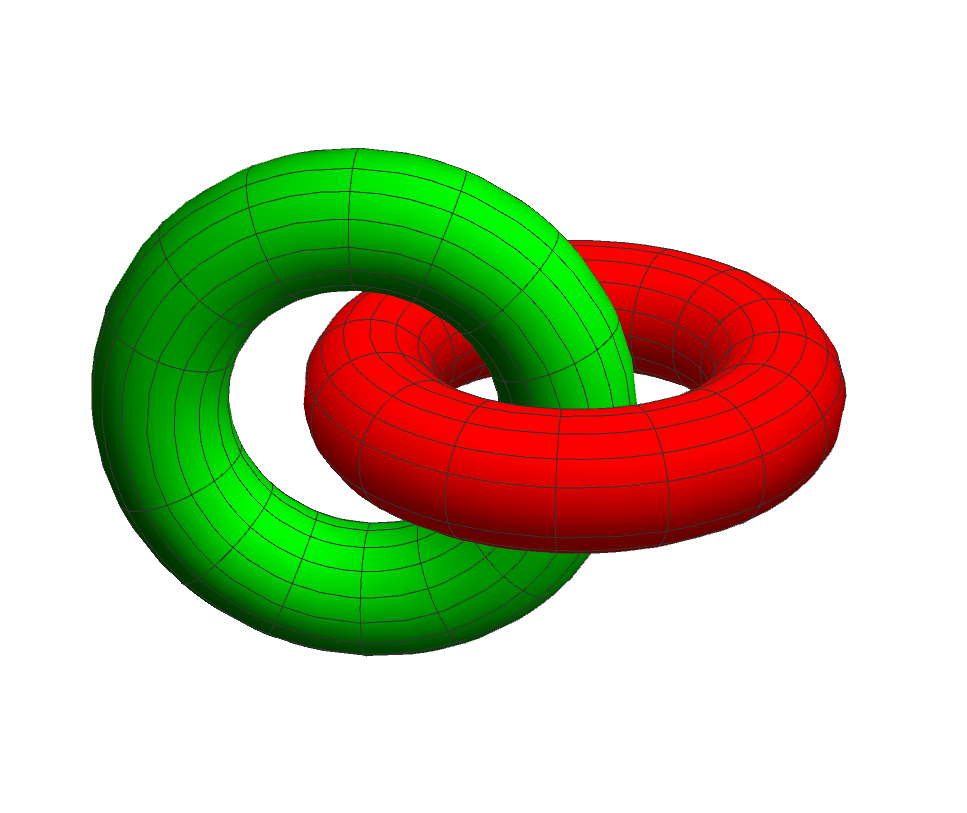}
      \caption{Pictorial diagram of $R_1$ (red) and $R_2$ (green). }
      \label{fig:linkedtubes}
    \end{center}
\end{figure}

The magnetic helicity associated with two flux tubes, with linking number $\mathcal{L}$, identical individual internal twists $\mathcal{T}$ and magnetic fluxes $\Phi$ is
\begin{equation} 
\label{eq:rings}
    H_{L} = 2(\mathcal{L} + \mathcal{T})\Phi^2,
\end{equation}
following \cite{Berger1999}. A simple example of such linked rings, $R_1$ and $R_2$ can be parameterised as
\begin{align}
    R_{1}(r,\theta,\phi) &= (R\cos(\theta) + r\cos(\theta)\cos(\phi), \nonumber \\
    & R\sin(\theta) + r\sin(\theta)\cos(\phi), r\sin(\phi)),
\end{align}
and
\begin{align}
&R_{2}(r,\theta,\phi)= (C_x, C_y, C_z) + (R\cos(\theta)+ r\cos(\theta)\cos(\phi),  \nonumber \\ 
&- r\sin(\phi), R\sin(\theta) + r\sin(\theta)\cos(\phi)),
\end{align}
for major radius $R$, minor radius $r \in[0,r_m]$, toroidal angle $\theta$ and poloidal angle $\phi$. The set ${C_x, C_y, C_z}$ denote the centre of $R_2$. An example with $r_m=0.3$ and $R=1$ is shown in Figure \ref{fig:linkedtubes}. We define the magnetic fields ${\bf B}_{R_i}$ of each ring as the sum of toroidal  ${\bf B}_{R_i t}$ and poloidal  ${\bf B}_{R_i p}$ components, with
\begin{align}
    \textbf{B}_{R_1, t} (x,y,z) &= B_{0} \bigg( -\frac{y}{\sqrt{(x^2 + y^2)}},  -\frac{x}{\sqrt{q_{1}}}, 0 \bigg),\\
 \nonumber    \textbf{B}_{R_1, p} (x,y,z) &=  \mathcal{T}B_{0} \bigg( \frac{xz}{r_{xy}}, -\frac{yz}{q_{1}}, 1-\frac{R}{\sqrt{q_{1}}} \bigg),\\
\nonumber    \textbf{B}_{R_2, t} (x,y,z) &= B_{0} \bigg( -\frac{z}{\sqrt{q_{2}}}, 0, -\frac{x + 1}{\sqrt{q_{2}}} \bigg),\\
  \nonumber  \textbf{B}_{R_2, p} (x,y,z) &= \mathcal{T}B_{0} \bigg( \frac{(x+1)y}{q_{2}}, -1+\frac{R}{\sqrt{q_{2}}}, \frac{yz}{q_{2}} \bigg),
\end{align} 
where $q_1 = (x^2 + y^2)$ and $q_2 = ((x+1)^2 + z^2)$.

\begin{figure}[h]
	\begin{center}
      \includegraphics[width=0.5\textwidth]{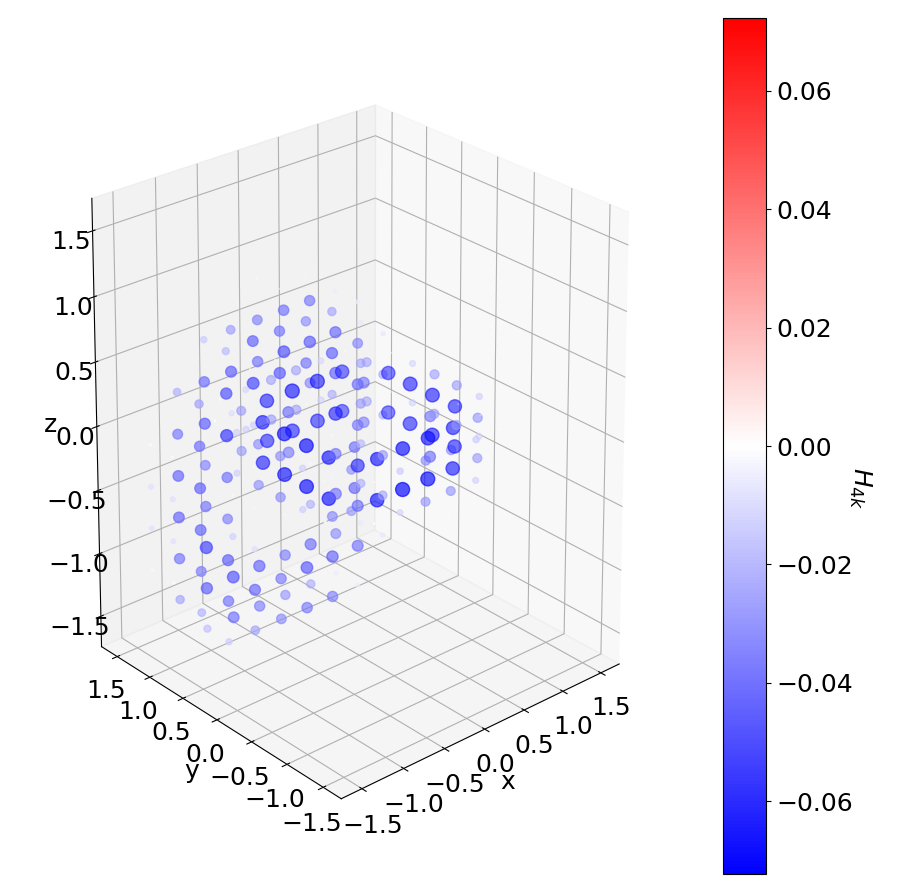}
      \caption{$H_{4k}$ for linked tubes with $\mathcal{T} = -5$. }
      \label{fig:linkedtubesscale4}
 	\end{center}
\end{figure}

We choose $R=1$ and $C_x = 1$, $C_y = C_z = 0$. Such an arrangement has an associated linking number of $\mathcal{L} = 1$, and we assign $\mathcal{T} = -5$, $B_0 = 7$ and $r_m= 0.3$, giving total magnetic helicity
\begin{equation}
    H_{\mathcal{L}=1, \mathcal{T} = -5} = (2 - 10)  \Phi^2 = -31.3,
\end{equation}
where $\Phi = 1.98$. In Figure \ref{fig:linkedtubesscale4}, we plot the magnetic helicity coefficients $H_{4k}$. The bubblegram  indicates a distribution of magnetic helicity in correspondence to the distribution of the magnetic fields themselves, which we can attribute the the magnetic twist.


\begin{figure*}[h]
\subfloat[Scale $s=1$]{\includegraphics[width=0.3\textwidth]{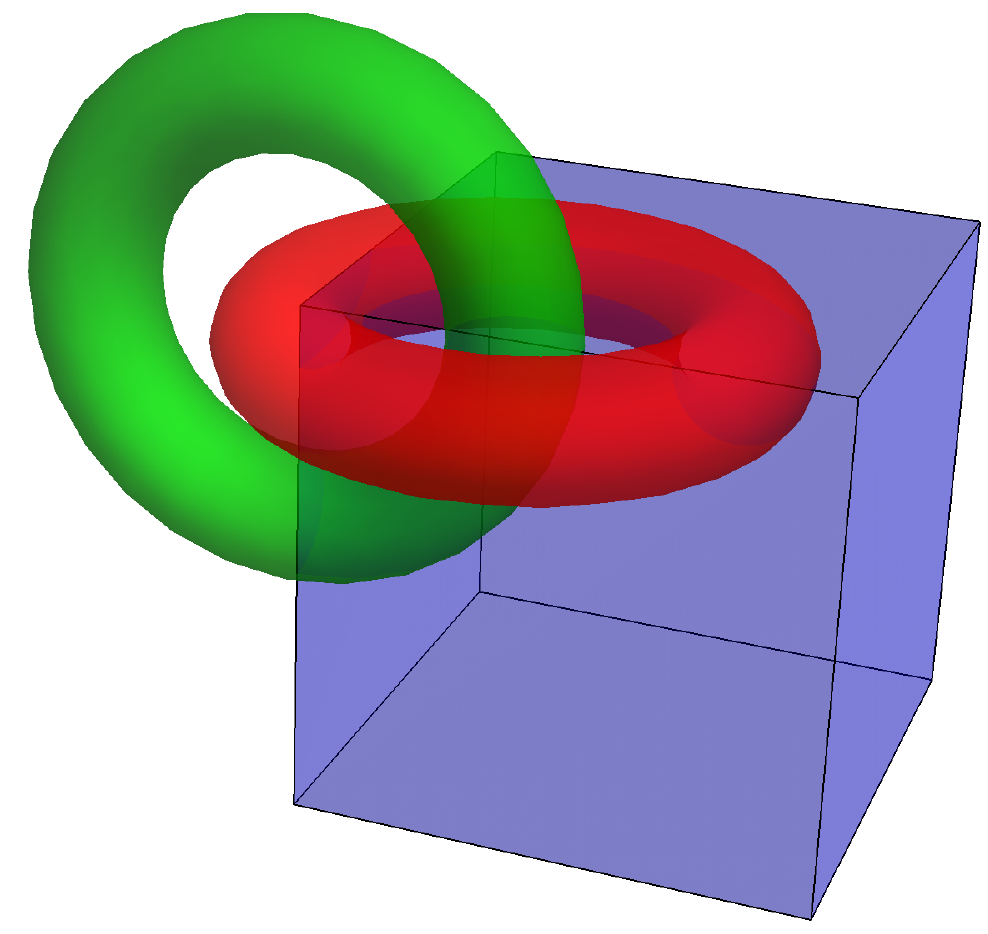}}\quad \subfloat[Scale $s=2$]{\includegraphics[width=0.3\textwidth]{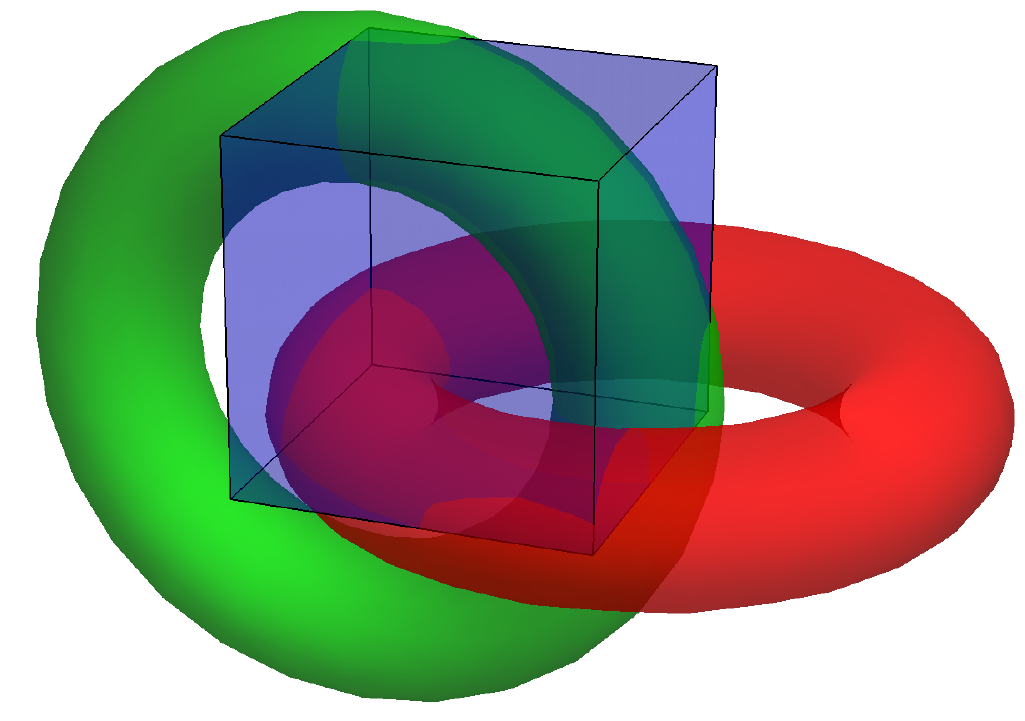}}\quad \subfloat[Scale $s=3$]{\includegraphics[width=0.3\textwidth]{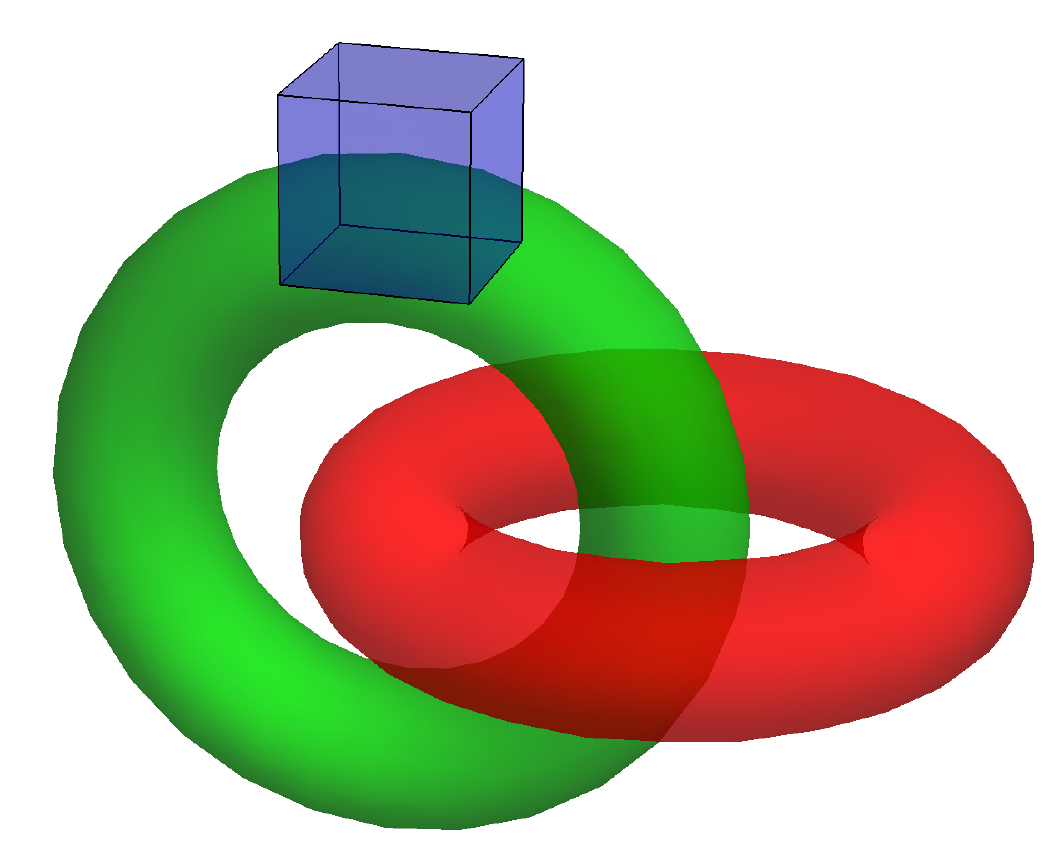}}
\caption{\label{scaleboxes} Figures indicating the contributing points of the density ${\bf A}\cdot{\bf B}$ at  various scales for a Harr wavelet decomposition. (a) $s=1$ and (b) $s=2$ the overlap of the two tubes in the region of compact support is clear. (c) scale $=3$ the region of compact support will generally only cover one tube.}
\end{figure*}

\begin{figure}[h]
\begin{center}
    \includegraphics[width=0.5\textwidth]{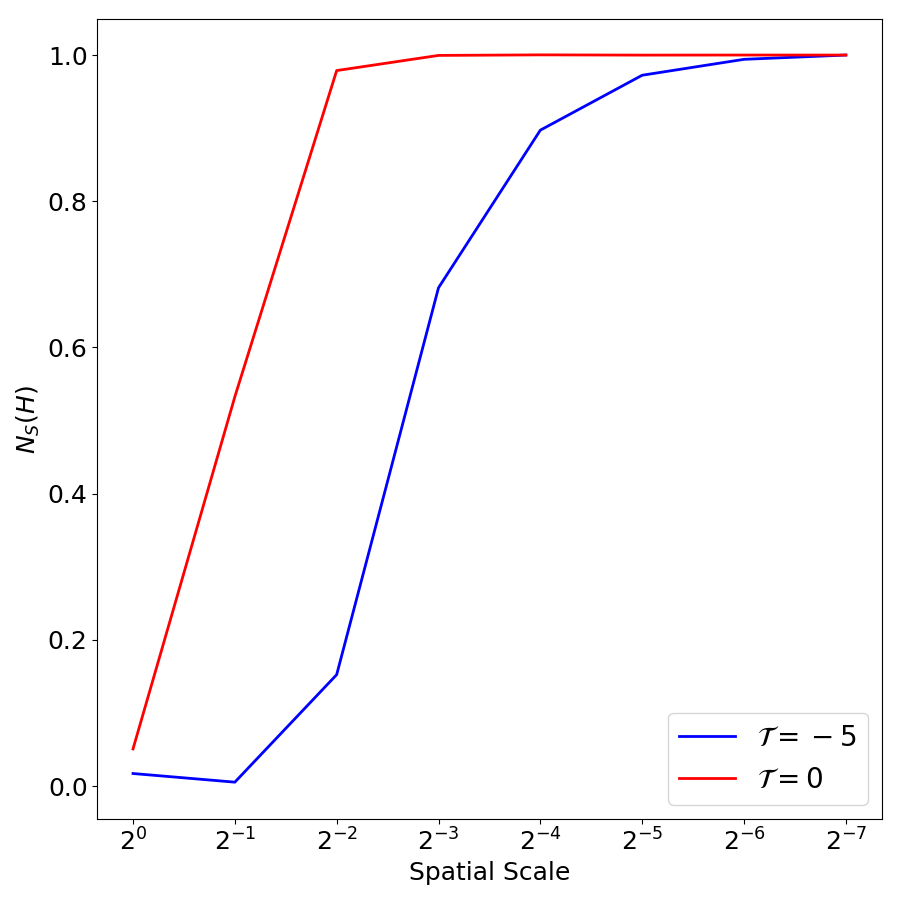}
      \caption{ $N_s(H)$ calculated for the two linked flux tubes, with either $\mathcal{T} = -5$ (red) or without ($\mathcal{T} = 0$) internal twist (blue). }
    \label{fig:linkedtubestotal}
\end{center}
\end{figure}

In Figure \ref{fig:linkedtubestotal}, we calculate the ratio of the multiresolution expansion of helicity with that of the analytical result, which we define by the measure
\begin{equation}
N_s(H) = \frac{\sum_{s' = 0}^{s} C_{s}(H)
}{H_L}
\end{equation} 
 described above (with and without internal magnetic twist). There is a clear distinction between the scales at which the ratio gets significantly close to $1$, as we would expect from the qualitative spatial separation between twist and linking. We see in Figure \ref{scaleboxes} the regions of compact support for the Harr wavelet's at scales $s=1,2$. The $s=1$ and $s=2$ scales tend to cover both tubes to some degree  whilst the scales $s=3$ and higher generally only cover one tube. This is reflected in Figure \ref{fig:linkedtubestotal} where we see the ${\cal T}=0$ field is dominated by scales $s=1,2$, as scales $s=3$ and higher will reflect that on the single tube interior scale there is no complex topology. By contrast the  ${\cal T}=5$ case has a more balanced distribution across the scales. 

\section{Helicity, Energy and Topology} \label{sec:helicityenergy}
We insert the full three-dimensional mutliresolution decomposition of the field ${\bf B}$  into the correlation function ${\bf C}$  (\ref{corrfunction}) to obtain,
\begin{align}
\nonumber {\bf C}({\bf x}) &= \frac{1}{2\pi}\sum_{s=0}^{s_m} \sum_{k=0}^{2^{s}-1}{\bf B}_{s'k'}\times\int_{V_z}\frac{{\bf r}}{r^2}\psi_{s'k'}(x',y',z) \rmd x'\rmd y'\\
& \quad {\bf r} = (x-x',y-y',0),
\label{windingdecomp}
\end{align}
where the parameter dependence of the wavelet function $\psi$ indicates the integration is over only the in-plane functions of the $3$-D wavelets. In order to compare the helicity to the energy (the helicity has an extra dimension of length) we note that if the planes $V_z$ have $x$ and $y$ widths $L$ and $a L$ respectively then we can write  $x = uL$ and $y = avL$ ($0 \leq u,v \leq 1$) so that ${\bf r_L} = L(u-u',a(v-v'),0)$, then 
\begin{align}
\label{windingdecompscaled}
\nonumber {\bf C}({\bf x}) &= \frac{L}{2\pi}\sum_{s=0}^{s_m} \sum_{k=0}^{2^{s}-1}{\bf B}_{s'k'} \\
& \times\int_{U_z}\frac{a\bf{r}_L}{r_L^2}\psi_{s'k'}(u',v',z)\rmd u'\rmd v',
\end{align}
where $U_z$ is the unit square. Inserting this into (\ref{helnogauge}) we obtain the helicity in terms of the multi-resolution expansion of the field ${\bf B}$ alone.  This calculation is most parsimoniously represented as a quadratic form  so we introduce some notation. We assume a Cartesian domain $U_z\times [0,h]$, with $U_z$ a unit plane at height $z$, then the following quantities are dependent only upon the chosen wavelet, not the magnetic field itself.
\begin{align}
\nonumber W_{i}^{s's k'k} = \frac{L}{2\pi} \int_{0}^{h}&\int_{V_z\times U_z}\frac{a r^i_{L}}{r_L^2}\psi_{s^{'}k^{'}}(u',v',z) \\
&  \psi_{sk}(x,y,z) \rmd u'\rmd v' \rmd x\rmd y\rmd z,
\label{wcoeff}
\end{align}
The cross-product in (\ref{windingdecomp})  can be represented using a skew-symmetric matrix $\textbf{M}^{s'sk'k}_{ij}$ which takes the form
\begin{equation}
\textbf{M}^{s's k'k}_{ij}  = \left(
\begin{array}{ccc}
 0 & 0 & -W_2^{s'sk'k} \\
 0 & 0 & W_1^{s'sk'k} \\
 W_2^{s'sk'k} &-W_1^{s'sk'k} & 0 \\
\end{array}
\right).
\end{equation}
Then, using the Einstein summation convention we have
\begin{equation}
H = \int_{V}{\bf C}\cdot{\bf B}\rmd V = \textbf{M}^{ss'k'k}_{ij}B^{j}_{s'k'}B^{i}_{sk}.
\end{equation}
We note that in general the wavelet orthogonality relationships cannot be applied to (\ref{wcoeff}) as the in-plane integrals are over different copies of $U_z$, however, the $z$ integration is over the same domain so $W_{i}$ will vanish if $n'\neq n$ (from the vectors $k=l'm'n'$ and $k=lmn$).

\subsection{Helicity as a skew symmetric operator}
We note that the helicity is being represented as product of the field at differing positions and scales through a skew-symmetric operator ${\bf M}$.  This is analogous to the result that the helicity in periodic domains can be represented as the skew symmetric part of the Fourier transform, as discussed in section \ref{fouriertwopoint}. In this case we use the decomposition ${\bf M}_{ij}^{s'sk'k} = L{\bf I}_{ij}^{s'sk'k}+ {\bf O}_{ij}^{s'sk'k}$, where ${\bf I}_{ij}^{s'sk'k}$ (the superscript labelling is for notational convince) is the identity matrix  (one such matrix for each $ss'kk'$) and 
\begin{equation}
{\bf O}^{s'sk'k} = \left(
\begin{array}{ccc}
-1 & 0 & -W_2^{s'sk'k} \\
 0 & -1 & W_1^{s'sk'k} \\
 W_2^{s'sk'k} & -W_1^{s'sk'k} & -1 \\
\end{array}
\right),
\end{equation}
so that 
\begin{equation} \label{eq:energytopology1}
H = L\textbf{I}_{ij}^{s'sk'k}B^{*j}_{s'k'}B^{i}_{sk} +  \textbf{O}_{ij}^{s'sk'k}B^{*j}_{s'k'}B^{i}_{sk}.
\end{equation}
The sum of contributions to the first term for which $(s',k')=(s,k)$ give the energy of the field (the multiresolution approximation to the energy to be precise), and we can thus decompose the sum as follows
\begin{align}
H &= L E  + N,\\
N &=L\textbf{I}_{ij}(1-\delta^{s's}\delta^{k'k})B^{*j}_{s'k'}B^{i}_{sk} + \textbf{O}_{ij}^{s'sk'k}B^{*j}_{s'k'}B^{i}_{sk}
\end{align}
where $\delta^{s's}$ is the Kronecker delta function. The operator $N$ contains additional topological information which constitutes the helicity. In the limit which the maximum scale parameter $s_m$  (\textit{i.e.} the smallest spatial scale) tends to $\infty$ this relationship is exact so there is a linear sum
\begin{equation}
H({\bf B}) = LE({\bf B}) +N({\bf B}),
\end{equation}
where $N$ is the multiresolution representation of a functional of the field which contains the topological information through the quantities $W_{i}^{s'sk'k}$.

\subsection{Helicity preserving field evolution}
  A field evolution for which $H$ is conserved requires that
\begin{equation}
\label{ratebalance}
\deriv{E}{t}=-\frac{1}{L}\deriv{N}{t}
\end{equation}
which would (approximately) apply in significantly low plasma $\beta$ resistive MHD simulations. 

\subsection{Field line helicity}
Using (\ref{helnogauge}) and (\ref{windingdecompscaled}) the fieldline helcity of a field  line $\gamma$ at scale $s$ and position $k=lm$ can be written as 
\begin{align} 
\mathcal{A}(\gamma) &=  \frac{L}{2\pi}\sum_{s=0}^{s_m} \sum_{k=0}^{2^{s}-1}\int_{\gamma}\frac{{\bf B}}{{\bf |B|}}\cdot{\bf B}_{s'k'}(z(s)) \nonumber \\
& \times\int_{U_z}\frac{a\bf{r}_L}{r_L^2}\psi_{s'k'}(u', v',z)\rmd u'\rmd v' \rmd{s},
\label{fieldlineheldecomp}
\end{align}
\textbf{where the summation over $k$ implies a 2-D multiresolution decomposition}, this is why the coefficient ${\bf B}_{s'k'}(z)$ of the multiresolution expansion has $z$ dependence.

Under ideal evolutions $\mathcal{A}(\gamma)$ is preserved so the sum of $\mathcal{A}_{s}(\gamma)$ must be preserved and changes in $\mathcal{A}_{s}(\gamma)$ must be balanced across the scales.
\subsection{Helicity preserving field evolution}
 A particular class of fields of significant interest in the solar physics community are braided fields for which $B_z>0\,\forall {\bf x} \in V$ and hence all field lines  pass through the domain from the bottom to top boundary. In such cases each field lines $\gamma$ can represented by the points ${\bf x}_0\in V_0$ where they are rooted, such that $\mathcal{A}_{sk}(\gamma)\equiv\mathcal{A}_{sk}({\bf x_0})$ and
\begin{align}
\label{fieldlinehelsum}
\nonumber H({\bf B})&=\int_{V_0}\mathcal{A}({\bf x_0})\rmd x \rmd y \\
& = \sum_{s=0}^{s_m}\sum_{k=0}^{2^s-1}\int_{V_0}\mathcal{A}_{sk}({\bf x_0})\rmd x \rmd y .
\end{align}
If the evolution is not ideal  but such that the helicity is conserved (low plasma $\beta$ resistive relaxations) the distribution $\mathcal{A}(\gamma)$ changes but the summation (\ref{fieldlinehelsum}) must be preserved. In particular we have an alternative means of calculating the value of the operator $N({\bf B})$.
\begin{align}
N({\bf B}) &= H({\bf B}) - LE({\bf B}) \nonumber \\
&= \sum_{s=0}^{s_m}\sum_{k=0}^{2^s-1}\int_{V_0}\mathcal{A}_{sk}({\bf x_0})\rmd x \rmd y- LE.
\end{align}
The advantage is that the field line helicity representation of $N$ is linear in both $s$ and $k$ so, for example, we can decompose the contributions to $N$ as the difference $H-LE$ at each scale $s$, and this decomposition is orthogonal. It is this form which we choose to utilize in this study.

\section{Fieldline Helicity Example: Analytical Magnetic Reconnection}
\begin{figure}
\subfloat[]{\includegraphics[width=8cm]{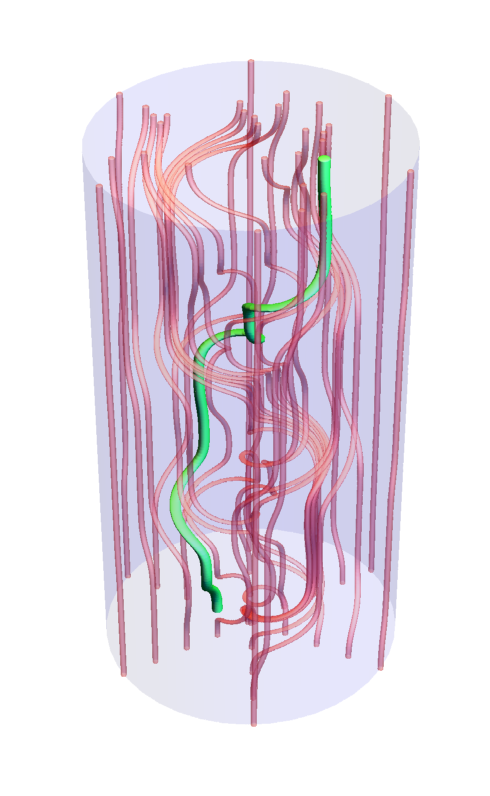}}\quad \subfloat[]{\includegraphics[width=8cm]{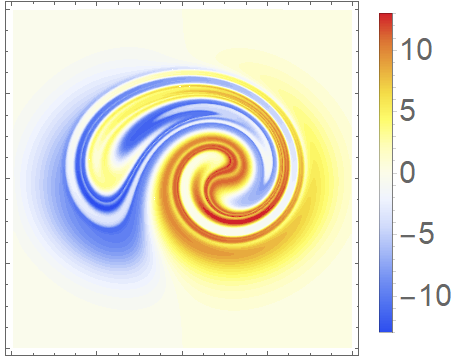}}
\caption{\label{dundeebraidflh}Figures indicating the entangled geometry of the braided field (\ref{braidfieldtube}). (a) indicates a subset of the field lines in the region where the fields opposing twist units overlap. The field line helicity of the green field line indicated would have contributions due to its own complex geometry as well as its entanglement with the field. (b) the field line helicity distribution (calculated using the code used in \cite{prior2018quantifying}) of (\ref{eq:dundeereconeq}) with $t=0$, there is significant small scale structure indicating he field's complex entanglement.}
\end{figure}
\label{sec:flhel}
Following the the resistive MHD based braiding experiments in \cite{wilmot2009magnetic,wilmot2011heating,russell2015evolution} we define a field  composed of exponential twist units ${\boldsymbol B}_t(b_0,k,a,l,x_c,y_c,z_c)$ given by
\begin{align}
&\nonumber {\bf B}_t(b_0,k,a,l,x_c,y_{c},z_{c}) =\nonumber \\ 
&\frac{2 b_0 k}{a}\mathrm{exp}\bigg(-\frac{(x-x_{c})^2+(y-y_{c})^2}{a^2}- \frac{(z-z_c)^2}{l^2}\bigg) {\bf R},
\label{bt}
\end{align}
where
\begin{align}
 {\boldsymbol R}&= (-(y-y_c),x-x_c,0).
\end{align}
The parameter $b_0$ determines the strength of the field, $a$ the horizontal width of the twist zones, $l$ their vertical extent and $k$ the handedness of the twist ($k=1$ is right handed). The centre of rotation is $(x_{c},y_{c},z_{c})$. The braided field is then defined as a superposition of $n$ pairs of positive and negative twists and a uniform vertical background field 
\begin{align}
&{\bf B}_{b}(1,a,l,d,z_0,s_d,n) = \nonumber \\
&\sum_{i=1}^{n}\bigg[ {\bf B}_t(1,1,a,l,-d,0,z_0+s_d(i-1)) \nonumber \\
 &+ {\bf B}_t(1,-1,a,l,0,d,0,z_0+s_d(i-1))\bigg]  + \hat{z},
 \label{braidfieldtube}
\end{align}
where, $d$ is the offset from the central axis, and $s_d$ is the vertical spacing between consecutive twists (of  the same sign) and $z_0$ the height of the first twist unit. By altering the extent of the twist units (the parameters $a$ and $l$) one can control the overlap the twist units. The field lines in the region of overlap show significant entanglement (Figure \ref{dundeebraidflh}(a)) a property very well captured by the field line helicity distribution $\mathcal{A}(\gamma)$ (Figure \ref{dundeebraidflh}(b)). The helicity of this field is (with a suitable choice of parameters) essentially zero owing to the balance of positive and negative twisting. It was found that under a high Reynold's number resistive MHD relaxation, under which the helicity is approximately conserved \cite{wilmot2011heating,russell2015evolution}, that the field was able to simplify via localised reconnection into (roughly) a pair of oppositely twisted flux ropes.

\begin{figure*}
\begin{center}
    \includegraphics[width=1\textwidth]{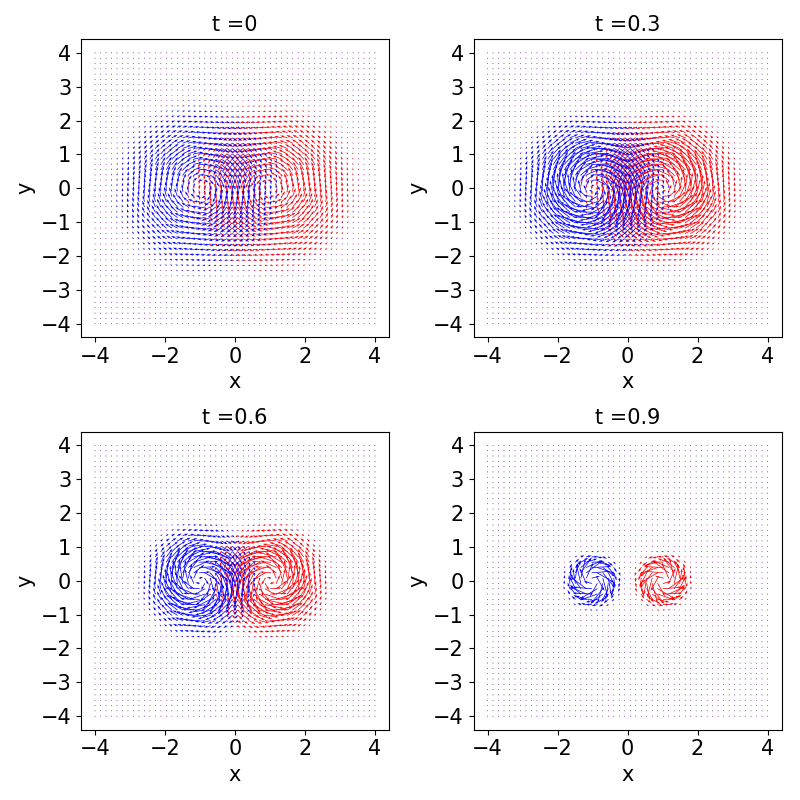}
      \caption{Vector plot at four time steps $t = 0, 0.3, 0.6,0.9$ at $z = 0$ of the magnetic field given by equation \ref{eq:dundeereconeq}. Red (blue) denotes the positively (negatively) twisted regions.}
    \label{fig:dundeevecplot}
\end{center}
\end{figure*}

\begin{figure*}[h] 
  \begin{center}
    \includegraphics[width=1\textwidth,clip=]{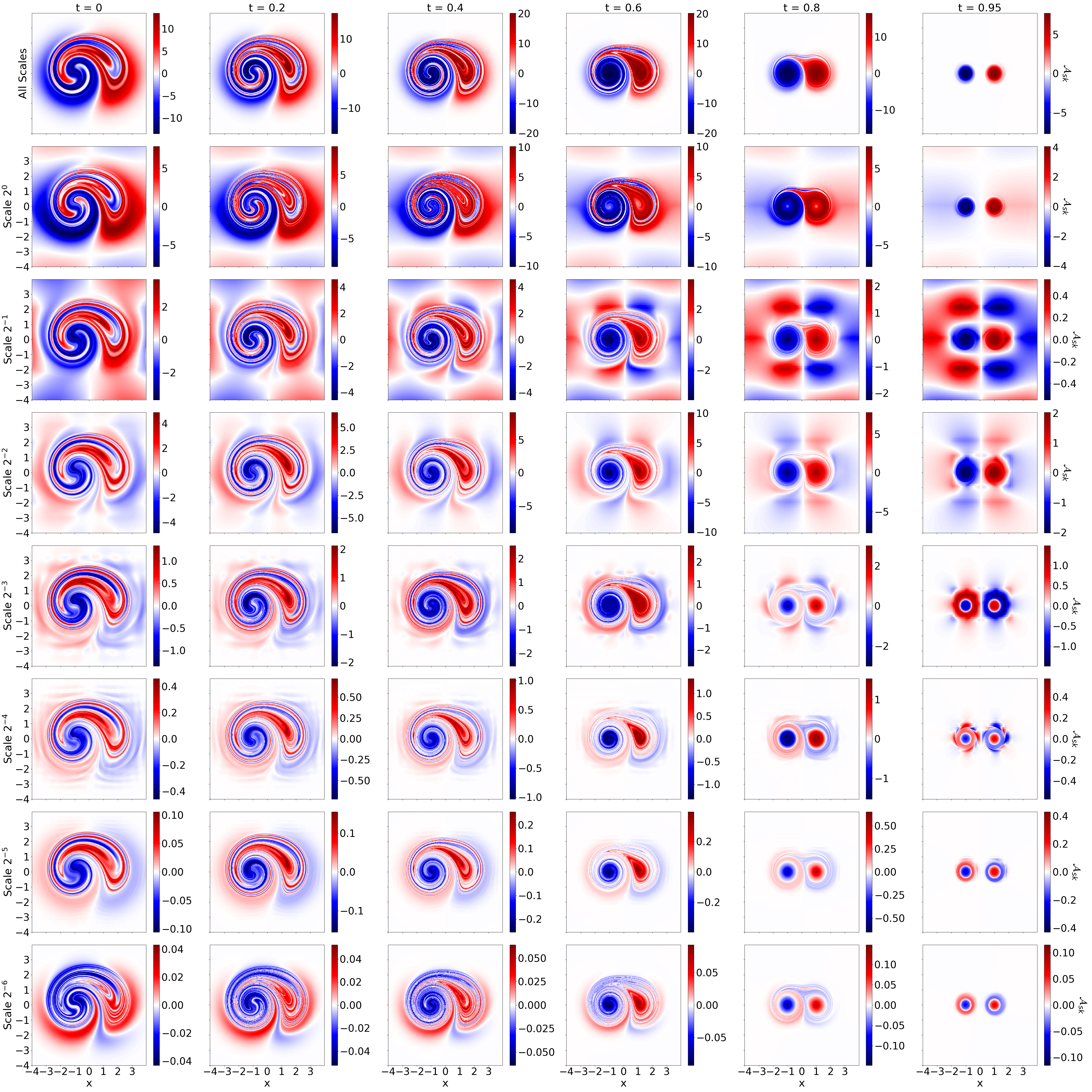}
  \end{center}
  \caption{Field line helicity distributions $C_s(\mathcal{A})({\bf x}_0)$ associated with time steps $t = 0, 0.2, 0.4, 0.6, 0.8, 0.95$ of an analytical reconnection simulator, in a domain $[-4, 4]^2$ in $x,y$ and $[-24 ,24]$ in $z$, with $400 \times 400$ field lines. The vertical direction indicates increasing $s$ (except for the top line which is the total sum over $s$ (it is the actual distribution $\mathcal{A}$). The horizontal direction indicate increasing $t$.}
\label{fig:B3scales}
\end{figure*}

To keep matters simple in this first application of the multiresolution decomposition $\mathcal{A}_{sk}$, we define a rough analytic approximation  of this relaxation process with the following parameterised magnetic field:
\begin{align}
\textbf{B}  = {\bf B}_{b}(1, D_1(t),D_2(t),1 ,-20,8, 3)
 \label{eq:dundeereconeq}
\end{align}
where
\begin{eqnarray*}
	D_1(t) &=& \sqrt{2(1-t)},\\
	D_2(t) &=& 2(1 + 2t),
\end{eqnarray*}
this field is considered in a domain $x,y\in[-4,4],\,z\in[-24,24]$, these are the dimensions (and parameters for $t=0$) used in  \cite{wilmot2009magnetic,wilmot2011heating,russell2015evolution}.
As $t$ increases the twisted units become more and more separated in the horizontal direction, as shown in Figure \ref{fig:dundeevecplot}. The twist units (with the same sign) also merge vertically to form two non overlapping twisted flux tubes at $t=1$. The decrease in overlap between the oppositely twisted units tends also reduces the complex field entanglement (as we shall shortly see this is not true for low $t$). It was checked numerically that the total helicity $H({\bf B},x,y,z,t)$ (essentially) remains zero for all $t$, a property designed to approximate the numerically observed conservation of helicity in the low plasma $\beta$ MHD simulations.
The Fourier expansion of the magnetic helicity of this field is zero throughout (even when an absolute magnitude sum is used). In Figure \ref{fig:B3scales}, we present the field line distribution of the field line helicity decomposition $C_s(\mathcal{A})({\bf x}_0)$, remembering that for the field line helicity there is one such summation for \textbf{each} point ${\bf x}_0$ (\textit{i.e.} each field line) - hence this is still a spatial distribution. The evolution of these distributions is shown at times $t = 0, 0.2, 0.4, 0.6, 0.8, 0.95$.

A couple of observations are worth making. 
\begin{enumerate}
\item{At $t=0$ all scales $C_s(\mathcal{A})({\bf x}_0)$ show (to varying degrees) the complex mixing pattern present in the full distribution. This is a result of the field line geometry (i.e. the geometry of the green curve in Figure \ref{dundeebraidflh}(a)). Eventually this pattern disappears as the field lines reconnect and disentangle, again this is true of all scales.}
\item{There is a a surrounding distribution which is most clear at the scales $s=1,2$, this persists throughout the relaxation. This is the twisted field structure of the field itslef, as indicated in the twisted tube example of section \ref{sec:opptwisttubes} twisted tube structures (which always compose the field in some manner) are dominated by contributions at these scales. Over the whole sum (over $s$ at each $t$) these contributions cancel.}
\end{enumerate}

%\begin{figure}[h] 
%  \begin{center}
%    \includegraphics[width=1.0\textwidth,clip=]{B3scales.png}
%  \end{center}
%  \caption{Field line helicity associated with time steps $t = 0, 0.2, 0.4, 0.6, 0.8, 0.95$ of an analytical reconnection simulator, in a domain $[-4, 4]^2$ in $x,y$ and $[-24 ,24]$ in $z$, with $400 \times 400$ field lines.}
%\label{fig:B3scales}
%\end{figure}
\begin{figure*} 
  \begin{center}
    \includegraphics[width=1.0\textwidth,clip=]{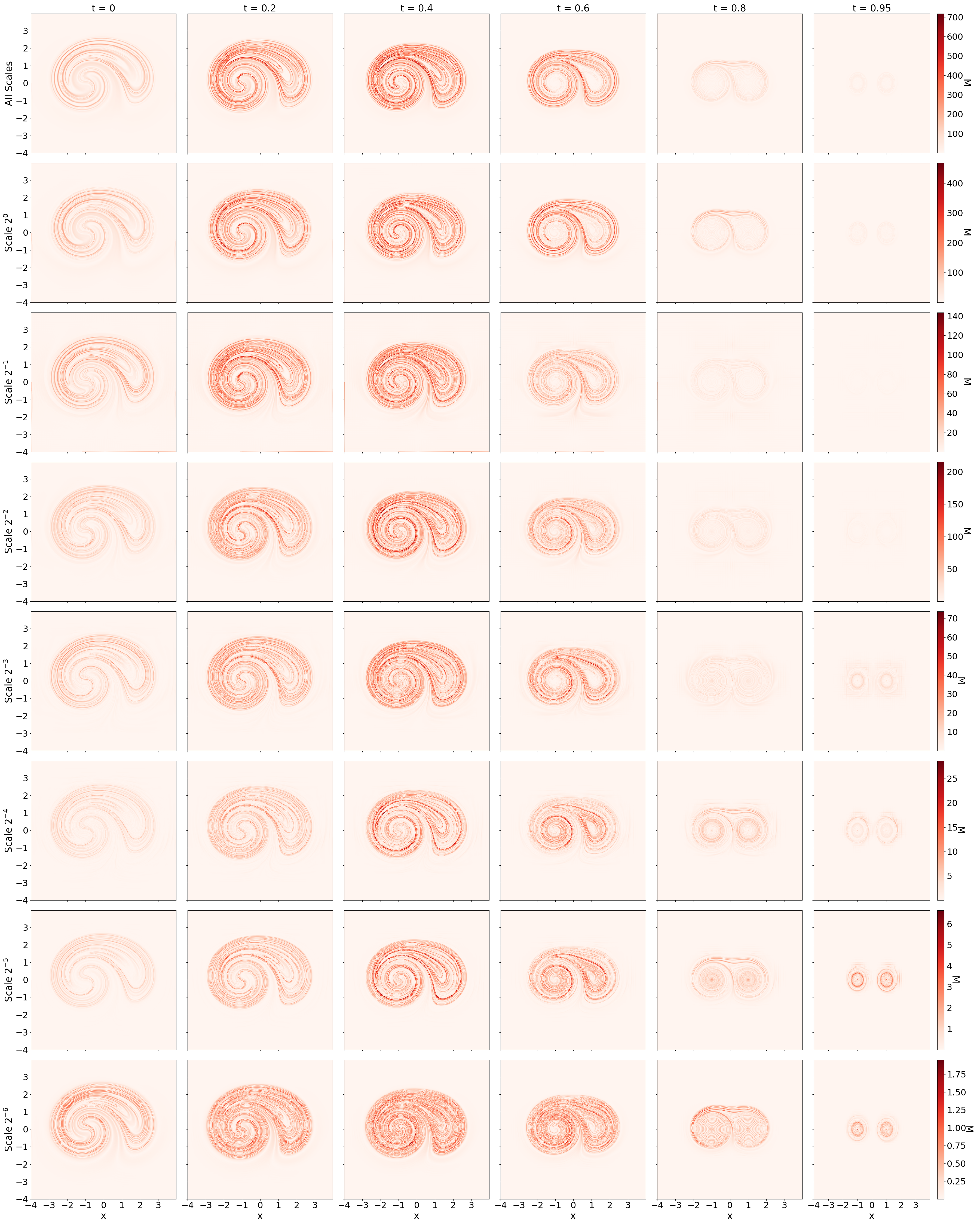}
  \end{center}
  \caption{Mixing M associated with time steps $t = 0, 0.2, 0.4, 0.6, 0.8, 0.95$ of an analytical reconnection field changing in time (\ref{eq:dundeereconeq}), in a domain $[-4, 4]^2$ in $x,y$ and $[-24 ,24]$ in $z$, with $400 \times 400$ field lines.}
\label{fig:B3mixing}
\end{figure*}

To quantify the entanglement variation highlighted in the first point  we define a mixing parameter $\rm{M}$ as
\begin{equation}
\rm{M} = |\nabla \mathcal{A}_{s}| = \bigg( \bigg|\frac{\partial C_s(\mathcal{A})({\bf x}_0)}{\partial x}\bigg|^2 +  \bigg|\frac{\partial C_s(\mathcal{A})({\bf x}_0)}{\partial y}\bigg|^2 \bigg)^{\frac{1}{2}},
\end{equation}
which will highlight the regions in which we see a rapid change in sign between positive and negative field line helicity $C_s(\mathcal{A})({\bf x}_0)$. Admittedly this will also capture simpler radial decay, but such contributions should be sufficiently weaker. The mixing associated with each scale, in the style of Figure \ref{fig:B3scales}, is shown in Figure \ref{fig:B3mixing}. There are two observations. First that the mixing actually increases at first up to $t=0.4$ then it decays. Second that the decay is more pronounced at larger length scales (smaller $s$).

\begin{figure*}[!h]
\begin{center}

    \begin{minipage}{0.45\textwidth}
      \includegraphics[width=\textwidth]{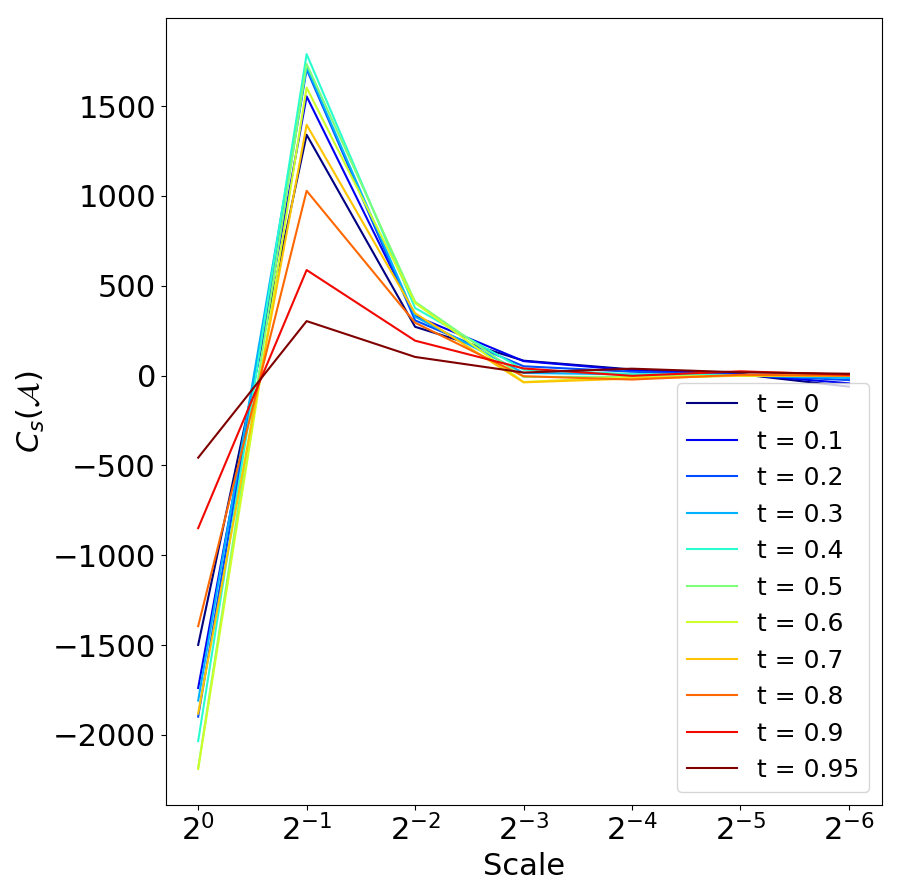}
      \caption{The total fieldline helicity power $C_{S}(\mathcal{A})$ attributed to each spatial scale, over time periods $t = 0$ to $t = 0.95$ for analytical reconnection via Dundee braids.}
      \label{fig:B3_scalepower}
    \end{minipage}\hfill
    \begin{minipage}{0.45\textwidth}
      \includegraphics[width=\textwidth]{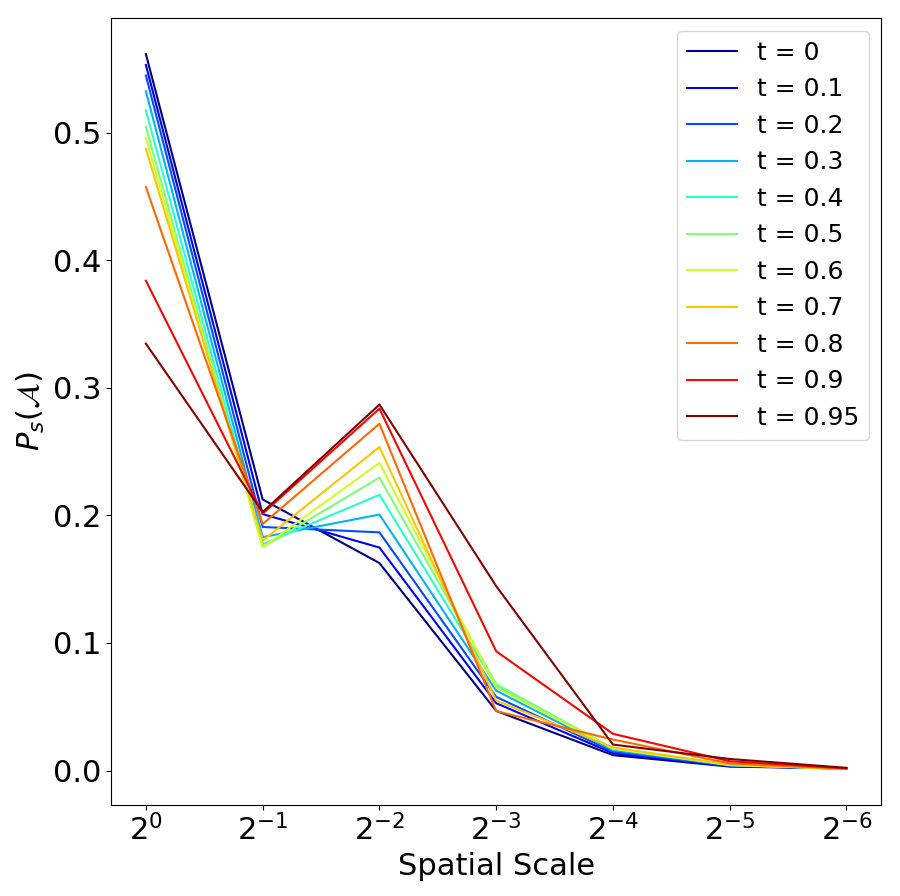}
      \caption{The total fieldline helicity absolute normalised power $P_{S}(\mathcal{A})$ attributed to each spatial scale, over time periods $t = 0$ to $t = 0.95$ for analytical reconnection via Dundee braids.}
      \label{fig:B3_abspower}
    \end{minipage}\hfill
    \\
    \begin{minipage}{0.45\textwidth}
      \includegraphics[width=\textwidth]{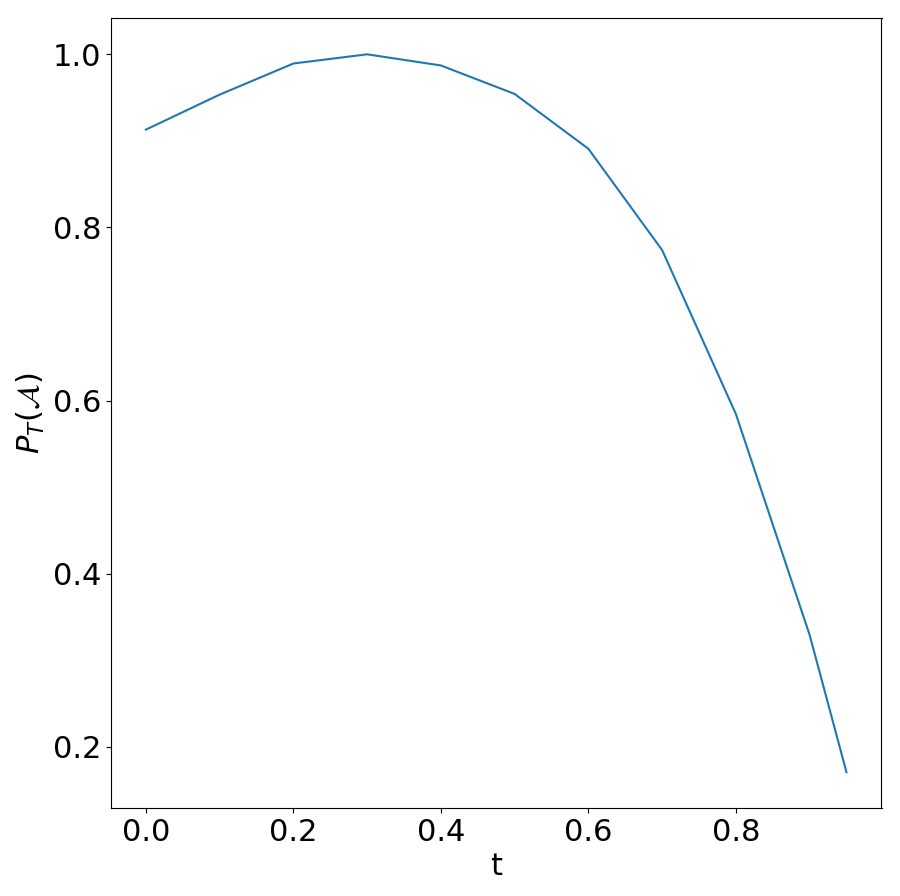}
      \caption{$P_T(\mathcal{A})$ from $t = 0$ to $t = 0.95$. }
      \label{fig:B3_abssumpower}
    \end{minipage}\hfill
    \begin{minipage}{0.45\textwidth}
      \includegraphics[width=\textwidth]{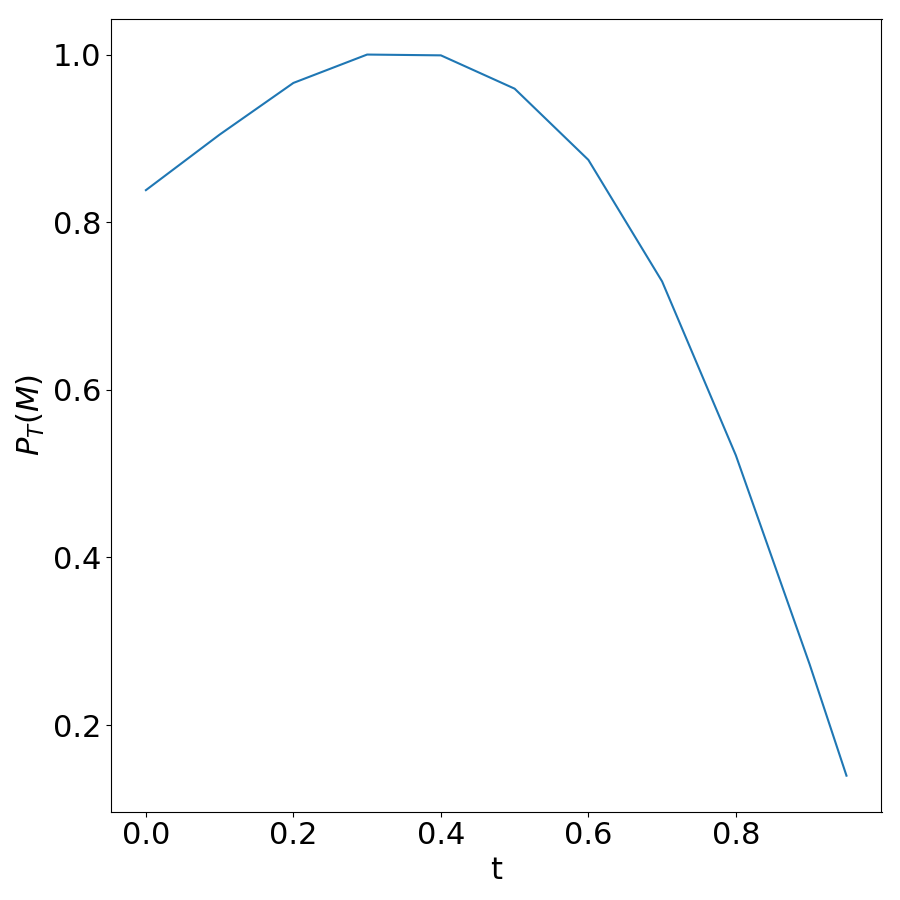}
      \caption{$P_T(M)$ from $t = 0$ to $t = 0.95$. }
      \label{fig:B3_absmixing}
    \end{minipage}\hfill
    
    \end{center}
\end{figure*}

In Figures \ref{fig:B3_scalepower} we plot the total signed contribution $C_s(\mathcal{A})$ (summed over the distributions shown in Figure \ref{fig:dundeevecplot}) as a function of scale for various $t$. There is always (approximately) as much negative as positive contribution, reflecting the total helicity conservation of the field. These values are dominated by the lower scale. Their relative magnitudes increase up to about $t=0.4$ then decrease over time. It is interesting that the balance of positive and negative values is always maintained by the same scales (albeit with decreasing magnitudes). In Figure \ref{fig:B3_abspower} we plot the absolute power $P_s(\mathcal{A})$ associated with each spatial scale for time steps $t = 0$ to $t = 0.95$. For early times the the values (mostly) decrease with $s$ (roughly following a power law). However, as the twisting units separate and merge the scale $s=2$ becomes more prevalent, reflecting the coherent development of the twisted flux ropes. In figures \ref{fig:B3_abssumpower} and \ref{fig:B3_absmixing} we see the total power normalised power across all scales of both the field line helicity $\mathcal{A}$ and the mixing $M$ as a function of time, given by
\begin{equation}
P_{T}(H) = \frac{\sum_{k=0}^{2^s-1}\vert H_{t,sk}\vert}{\rm{max}_t \sum_{k=0}^{2^s-1}\vert H_{t,sk} \vert }.
\end{equation}
 Qualitatively the plots are very similar, showing a peak around $0.35$ and then a relatively large drop as the twist units properly separate. The degree of mixing is determined by a ratio of the $z$ decay and the $x$-$y$ overlap: if the $z$-decay is too weak, the two regions of twist will cancel and reduce the degree of mixing (assuming the twist regions overlap).

%\begin{figure}[ht] 
%  \begin{center}
%    \includegraphics[width=1.0\textwidth,clip=]{B3mixing.png}
%  \end{center}
%  \caption{Mixing $M_s$ associated with time steps $t = 0, 0.2, 0.4, 0.6, 0.8, 0.95$ of an analytical reconnection simulator, in a domain $[-4, 4]^2$ in $x,y$ and $[-24 ,24]$ in $z$, with $400 \times 400$ field lines. Again the vertical direction indicates increasing $s$ and the horizontal direction $t$.}
%\label{fig:B3mixing}
%\end{figure}

\begin{figure*}[!h] 
  \begin{center}
    \includegraphics[width=1.0\textwidth,clip=]{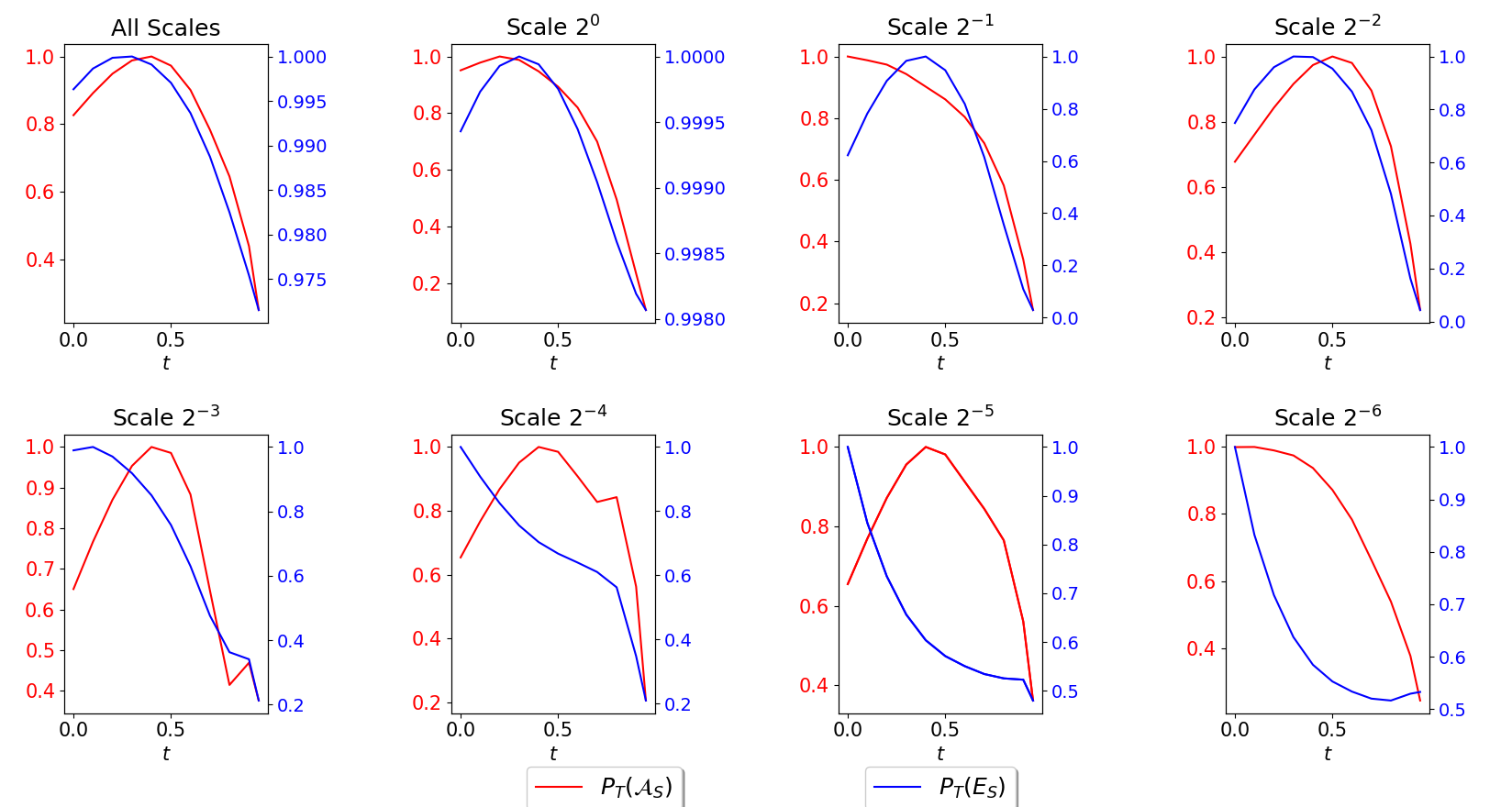}
  \end{center}
  \caption{{$P_T$ associated with the multiresolution analysis of fieldline helicity versus that attributed to the full three-dimensional decomposition of magnetic energy, where normalisation is scale dependent.}}
\label{fig:B3_energy_flh_comparison}
\end{figure*}

We can also directly compare the evolution of the scaled-fieldline helicity with that of magnetic energy, as shown in Figure \ref{fig:B3_energy_flh_comparison}, where we plot the absolute normalised magnetic energy against that of fieldline helicity, normalised within each scale ($P_T(\mathcal{A}_S)$ versus $P_T(E_S$).

The correlation between the two time series is seen to decrease as the spatial scale decreases in size. Their relative decay is most strongly aligned at scales $2^{0}-2^{-2}$. Whilst the decay associated with fieldline helicity power is fairly consistent at all scales, the decay of magnetic energy is opposite to that of field line helicity at scales $2^{-5}$ and $2^{-6}$.  It is no surprise that the scales $s=0,2$ are the most aligned, as we see in Figure  \ref{fig:B3_scalepower} these are the dominant contributors to the field line helicity variations in the field. As the magnitude of these peaks rise (up to $t=0.3$) and fall $t>0.3$ (Figure  \ref{fig:B3_scalepower}) so concurrently does the energy. This is a potentially important observation: that the variations in the the multi-resolution decomposition of the field line helicity $\mathcal{A}_{sk}$ are intimately correlated with the variations in energy in the field.  In future studies it would be interesting to see whether this correlation is maintained in resistive relaxation simulations.

Note that spatial scales $2^{-1}, 2^{-2}$ (in the $x-y$ plane) corresponds to the distribution of twisted regions, and as such we would expect it to contain a large portion of the power. Scale $2^{0}$ corresponds, for the fieldline helicity expansion, corresponds to the topological complexity of the line integrals themselves along each fieldline (there is only one-coefficient of $B_{sk}$ filling our whole domain). This relation between energy and topology was determined analytically in Section \ref{sec:wavelethelformula} for the winding gauge, and as such we should also hope to observe it in resistive relaxation simulations.

\section{Flux of Magnetic Helicity}
The flux of magnetic helicity through a surface is typically defined by 
\begin{align} \label{eq:helflux}
\frac{dH}{dt} &= -2\int_{V} \textbf{E} \cdot \textbf{B} \ \rm{d}^3 \rm{x}\nonumber \\ 
& + \int_{S} \big( (\textbf{A}_{P} \cdot \textbf{v}) \textbf{B}  + (\textbf{A}_{P} \cdot \textbf{B}) \textbf{v} \bigg)\cdot \hat{ \textbf{n}} \ \rm{d}^2 \rm{x},
\end{align}
for reference field $\textbf{A}_{P}$ uniquely defined by the appropriate boundary conditions of magnetic field $\textbf{B}$, and velocity field $\textbf{v}$. The first term refers to dissipation within the volume, which has been shown to have an effective time scale less than energy dissipation, and we thus disregard it. The second expression can be interpreted as the sum of two individual fluxes: the effect of twisting motions on the boundary, and secondly the movement of magnetic field through the boundary.

Wavelet analysis allows us to define a fourth measure of helicity flux, giving an indication of how helicity moves spatially within the volume. An intuitive example of this could be a study of a coronal loop expanding through a simulated region, for which the twist associated with the flux rope would be seen to move spatially. Multiresolution analysis measures helicity as a set of coefficients $H_{sk}$ attributed to a given scale and spatial domain (with compact support). We can then simply define
\begin{equation}
\frac{dH_{t,sk}}{dt} = \frac{H_{t,sk} - H_{t- \delta t, sk}}{\delta t}, 
\end{equation}
in the form of a finite difference approximation, for the multiresolution analyses of two adjacent time snapshots.

\begin{figure}
    \centering
    \includegraphics[width=0.5\textwidth]{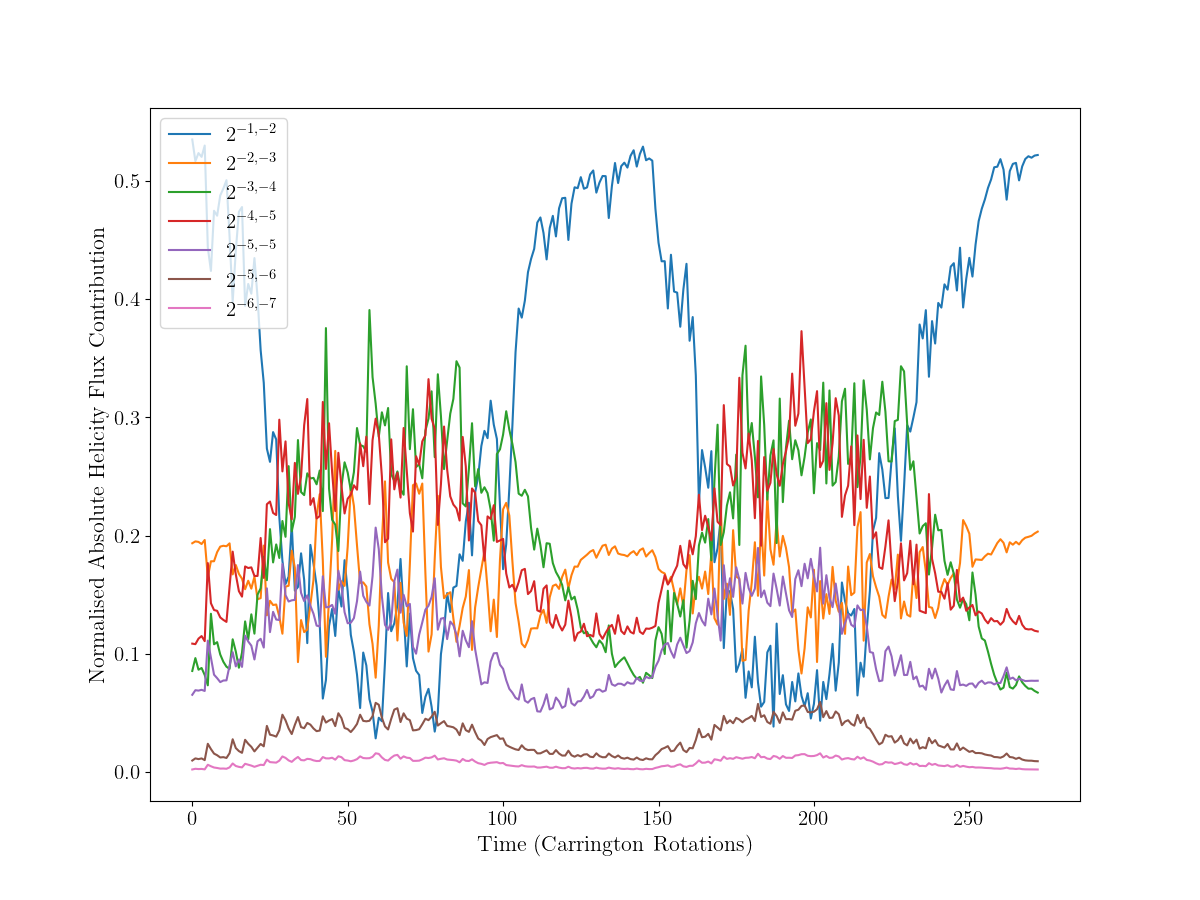}
    \caption{Helicity flux $dH/ dt _{sk}$ of equation (\ref{eq:mrahflux}) for a portion of the simulations of \cite{Jiang2011}, scale $2^{-a,b}$ refers to spatial scale $2^{-a}2$ in $\cos(\theta)$ and $2^{-b}2\pi$ in $\phi$. } 
    \label{fig:solarhflux}
\end{figure}

Further, we can perform a direct (2-D) multiresolution analysis on each term of the analytical measure of flux. For instance,
\begin{equation} \label{eq:mrahflux}
\frac{dH}{dt}_{sk} = \int_S (\textbf{A}_{P} \cdot \textbf{v}) \mathbf{\psi}_{sk} (\textbf{x}) \rm{d}^2 \rm{x}  \cdot \int_S  B_r(\textbf{x}) \mathbf{\psi}_{sk} \rm{d}^2 x,
\end{equation}
where we note that the z-spatial co-ordinate has been dropped again ($k = lm$). This is a multiresolution form of the helicity flux used in studies of the solar helicity flux through the hemisphere \cite{Hawkes2018}. Using the surface flux transport model simulations of \cite{Jiang2011}, we calculate the helicity flux associated with seven spatial scales in Figure \ref{fig:solarhflux}
This data covers their simulations for Solar Cycles $21$ and $22$, where time is counted from the beginning of cycle $21$. As each cycle develops, the helicity flux associated with the largest scale ($2^{-1, -2}$ in $(\cos(\theta), \phi)$, which equates to a hemispherical split), drops in line with an increase in helicity flux associated with $B_r$ of a smaller scale. This can be interpreted as the decreasing relative importance of polar (large scale) field relative to small-scale emerging active regions. This behaviour is seen to repeat over the course of two solar cycles (the end of the figure corresponds to the end of Cycle 22). 
\section{Conclusions}
We have demonstrated how a multiresolution decomposition can be applied to the magnetic helicity and field line helicity, crucial topological quantities in astrophysical applications of the MHD equations. This approach is compared to spectral helicity decompositions, which require periodic domains. The method of multi-resolution analysis has some significant advantages over this purely spectral approach:
\begin{enumerate}
\item{It requires no periodicity conditions on the domain thus has a far wider range of potential applications.}
\item{It yields information on the spatial decomposition of helicity in the field, this  is particularly useful for fields with significant heterogeneity of their entanglement.}
\end{enumerate}
On the first point the we have circumnavigated any issues  regarding gauge choice by instead using a concrete geometrical definition of helicity which combines the results of \cite{prior2014helicity} and \cite{berger2018generalized} to give a topologically meaningful definition of the helicity which does not depend only on the field ${\bf B}$ nor its vector potential. It has no requirements on the boundary conditions of the field to be valid. The second point is a direct consequence of decomposing the magnetic field ${\bf B}$ using a wavelet (multiresolution) expansion, rather than a Fourier series expansion.  The following explicit theoretical results were obtained.
\begin{enumerate}
\item{If a vector potential definition of the helicity is used (the winding gauge of \cite{prior2014helicity} is recommended to be consistent with the rest of the results), then the helicity can be written as a sum of the components of the multiresolution expansions of the field ${\bf B}$ and the vector potential {\bf A}, this is given by formula (\ref{helsum}). If the winding gauge is used then we have given an explicit geometrical interpretation of the coefficients $H_{sk}$ (at scale $s$ and position vector $k$) as indicated visually in Figure \ref{fieldlinehel}(a).  We demonstrate the efficacy of this method with the mutliresolution analysis correctly identifying the opposing twisting two flux tubes in (\ref{sec:opptwisttubes}) (where the Fourier decomposition does not). In section \ref{sec:rings} we show there is a clear scale separation of twisting and writhing components of helicity of a pair of linked flux ropes.}
\item{By using a purely geometrical definition of the helicity we can show it is possible to express the helicity as a linear sum:
\begin{equation}
H({\bf B}) = LE({\bf B}) + N({\bf B})
\end{equation}
where the operator $N$ is a sum over various contributions to the total winding (entanglement) of the field from the various scales and spatial components of the multi-resolution expansion of  the field ${\bf B}$, and $L$ is the characteristic horizontal length scale of the domain. This can be seen as a significant extension of the two point field correlation Fourier energy/helicity decomposition applicable for fields in periodic domains (see \textit{e.g} \cite{brandenburg2017two}). This decomposition not only places no requirement on the boundary conditions of the field but gives information about the spatial distribution of contributions to this sum. 
 }\item{The field line helicity $\mathcal{A}(\gamma)$, the average entanglement of the field line $\gamma$ with the rest of the field, can be composed into both spatial and scale components using a multiresolution analysis (see equation (\ref{fieldlineheldecomp})). Under  an ideal evolution, when the distribution of field line helicity is conserved, this decomposition could be used to provide insight as to how the field's topology redistributes both spatially and across scales \textit{i.e.} flux ropes kinking/expanding or buoyantly rising through the convection zone of the sun. In this initial study we applied the field line helicity decomposition to an analytic representation of a resistively relaxing magnetic braid whose total helicity is conserved (mimicking well known numerical experiments of low plasma $\beta$ resistive MHD relaxation of the same magnetic braid configuration \citep{russell2015evolution}). In this case the spatially integrated sum of the field line helicity at each scale, which is equal to the helicity and hence conserved, indicated that the conservation was maintained by a varying balance of entanglement on scales which reflected the varying filed line entanglement and the twisted structure of the underlying magnetic field. It was also seen that the variance in these contributions strongly correlated with the variations in energy of the field during its relaxation.
}\item{We demonstrate how to apply this multiresolution decomposition to helicity fluxes through a planar boundary. An example application of this to a surface flux transport model over two solar cycles is used to indicate the analysis separates the varying contributions from the large-scale polar field and the smaller scale active region contributions to the cycle variation. 
}
\end{enumerate}

In addition to these results and findings we have developed a number of simple methods/quantities which can be used to draw conclusions from the expansions, such as the scale total and power coefficients $C_s$  and $P_s$, and the mixing measure $M$ used to interpret the varying degree of complexity of the field line helicity decompositions in section \ref{sec:flhel}. This can be a difficult task as the multiresolution decomposition presents a significant and potentially overwhelming amount of information.

\subsection{Future directions}
As indicated in the title this is part one of a two part study. The second part of the study will be to apply these techniques to Resistive MHD simulations. Based on the results of this study, we propose that the following lines of inquiry should be a priority.
\begin{enumerate}
\item{It is known that there is a clear relationship between the Fourier energy and helicity spectrum in homogeneously driven turbulence (\textit{e.g.} \cite{brandenburg2017two}). The question to answer is whether there is a similar relationship in highly heterogeneous systems for a multiresolution decomposition of the energy and helicity. The energy scale/field line helicity scale correlation, found in the analytically driven braid relaxation of section \ref{sec:flhel} offers some promise, but it should be investigated as to whether this same behaviour manifests in the resistive MHD relaxations of \cite{wilmot2009magnetic,wilmot2011heating,russell2015evolution}.}
\item{What information can be obtained from the helicity energy decomposition? In particular, under a field evolution which preserves helicity the product represented by the operator $N$ must oppose that of the energy. Further, $N$ contains the topological information of the field. Since this decomposition applies at each spatial point of a discretized field  an in-depth analysis of the transfer between these two quantities may be able to yield information as to how reconnection activity can lead to a field relaxing to force free equilibrium. Of particular interest will be simulations which do not follow the Taylor relaxation hypothesis (those which relax to a non linear force free equilibrium) as it implies the assumption that the helicity is the only topological quantity not destroyed during relaxation is not true in general.}
\item{Can the decomposition be used to identify relatively large spatial scale substructure in heterogeneous turbulence ? \textit{i.e.} partial flux rope type structures.}
\item{Does the decomposition, applied to flux transport types simulations or magnetogram data indicate anything about the variations in behaviour of solar cycles?}
\end{enumerate}

\begin{acknowledgements}
We thank the UK STFC for their funding under grants ST/N504063/1 (GH) and {\color{red}...}. This collaboration was facilitated through the NORDITA programme on Solar Helicities in Theory and Observations.
\end{acknowledgements}
\bibliographystyle{aabib}
\bibliography{WaveletBib}

\end{document}